\documentclass[a4paper]{article}
\usepackage{epsfig}
\begin{document}
\newcommand{\be}{\begin{equation}}
\newcommand{\ee}{\end{equation}}
\newcommand{\bea}{\begin{eqnarray}}
\newcommand{\eea}{\end{eqnarray}}
\newcommand{\nn}{\nonumber \\}
\newcommand{\ex}[1]{{\rm e}^{#1}}
\newcommand{\de}{{\rm d}}
\newcommand{\ie}{{\rm i}}
\newcommand{\dens}{\rho_{\rm d}}
\newcommand{\ct}{{\rm c}_0}
\newcommand{\st}{{\rm s}_0}
\newcommand{\frp}[2]{\frac{\partial #1}{\partial #2}}
\newcommand{\KK}[2]{K\left(#1\mbox{\scriptsize
$\begin{array}{cc}#2\end{array}$}\right)}
\newcommand{\w}[1]{{\bf #1}}
\newcommand{\Dphi}{D^{l}_{mm'}(\alpha,\beta,\phi)}
\newcommand{\Dzero}{D^{l}_{0m}(\alpha,\beta,0)}
\newcommand{\Dzeros}{D^{l}_{0m'}(\alpha,\beta,0)}
\newcounter{tabnr}
\newcommand{\assigncounter}[1]{\stepcounter{tabnr}\newcounter{#1}%
\setcounter{#1}{\value{tabnr}}}
%
% Uses:
% sphere.eps,
% spherehole.eps (generated by sphole.p and help.eps),
% tabcyclg, tabcyclu,
% tabted, tabtedc,
% taboct, taboctc,
% tabdod (all generated by fl3d3.p),
% tabeqshape
%

\vspace*{1cm}

\centerline{\huge \bf Floating Bodies of Equilibrium in}
\medskip

\centerline{\huge \bf Three Dimensions.}
\medskip

\centerline{\huge \bf The central symmetric case}

\vspace{1cm}

\begin{center}
\bf Franz Wegner, Institut f\"ur Theoretische Physik \\
Ruprecht-Karls-Universit\"at Heidelberg \\
Philosophenweg 19, D-69120 Heidelberg \\
Email: wegner@tphys.uni-heidelberg.de
\end{center}
\vspace{1cm}

\paragraph*{Abstract} Three-dimensional central symmetric bodies different from
spheres that can float in all orientations are considered. For relative density
$\rho=\frac 12$ there are solutions, if holes in the body are allowed.
For $\rho\not=\frac 12$ the body is deformed from a sphere. A set of nonlinear
shape-equations determines the shape in lowest order in the deformation. It is
shown that a large number of solutions exists. An expansion scheme is given,
which allows a formal expansion in the deformation to arbitrary order under the
assumption that apart from $x=0,\pm1$ there is no $x$, which obeys $P_{p,2}(x)=0$ for two different integer $p$s, where $P$ are Legendre functions.

\section{Introduction and Summary}

A long standing problem asked by Stanislaw Ulam in the Scottish Book
\cite{Scottish} (problem 19) is, whether a sphere is the only solid of uniform
density which will float in water in any position. Such a solid is called a
{\it floating body of equilibrium}. It will be in indifferent equilibrium in all orientations.

The simpler, two-dimensional, problem to find non circular cross-sections of a
long cylindrical log which floats without tending to rotate (the
axis of the log is assumed to be parallel to the water surface.) was solved for
relative density $\dens=1/2$ in 1938 by Auerbach \cite{Auerbach} and for
densities $\dens\not= 1/2$ by the present author
\cite{WegnerI,WegnerII,WegnerIII}.

Here we start to investigate the problem for three dimensional systems.
In section 2 the basic equations for a partly immersed body in
indifferent equilibrium are derived. In the following section the theorem that a
star-shaped inversion-symmetric body in arbitrary dimension $d$ and density
$1/2$ is a sphere, is reconsidered. One easily sees that a floating body of equilibrium, which is not star-shaped can have a shape different from a ball.
A class of such bodies is explicitly given.

The following sections, which are concerned with star-shaped floating bodies of equilibrium with arbitrary relative densities, can be read without having read section 3.
In section 4 and 5 we start considering the expansion of the surface of a
floating body of equilibrium around the sphere. To perform the expansion the distance of the surface is
measured from the center of gravity. It is expanded in spherical harmonics,
which are defined as the eigenfunctions of the Laplacean operator on the unit
sphere with eigenvalues $-l(l+1)$. For given $l$ the eigenfunctions are linear
combinations of $2l+1$ linearly independent functions.
\smallskip

\noindent
\parbox[b]{7cm}{\hspace{4mm}
In section 6 the equations are considered for the deformation in first order.
If the angle $\theta_0$ is a zero of the associated Legendre
function $P_{p,-2}(\cos\theta_0)=0$, then the spherical harmonics with $l=p$
contribute to the deformation.\\
This equation is fulfilled at $\theta_0=\pi/2$ for all odd $p$, which
corresponds to $\dens=1/2$. Thus one has to expect that at this special density
there is a large set of solutions.\\
In the following we assume that with the exception of $\cos\theta_0=0,\pm 1$ there is no $\theta_0$, which solves $P_{p,-2}(\cos\theta_0)=0$ for two different integer $p$.} \hfill
\parbox[b]{4.5cm}{
\epsfig{file=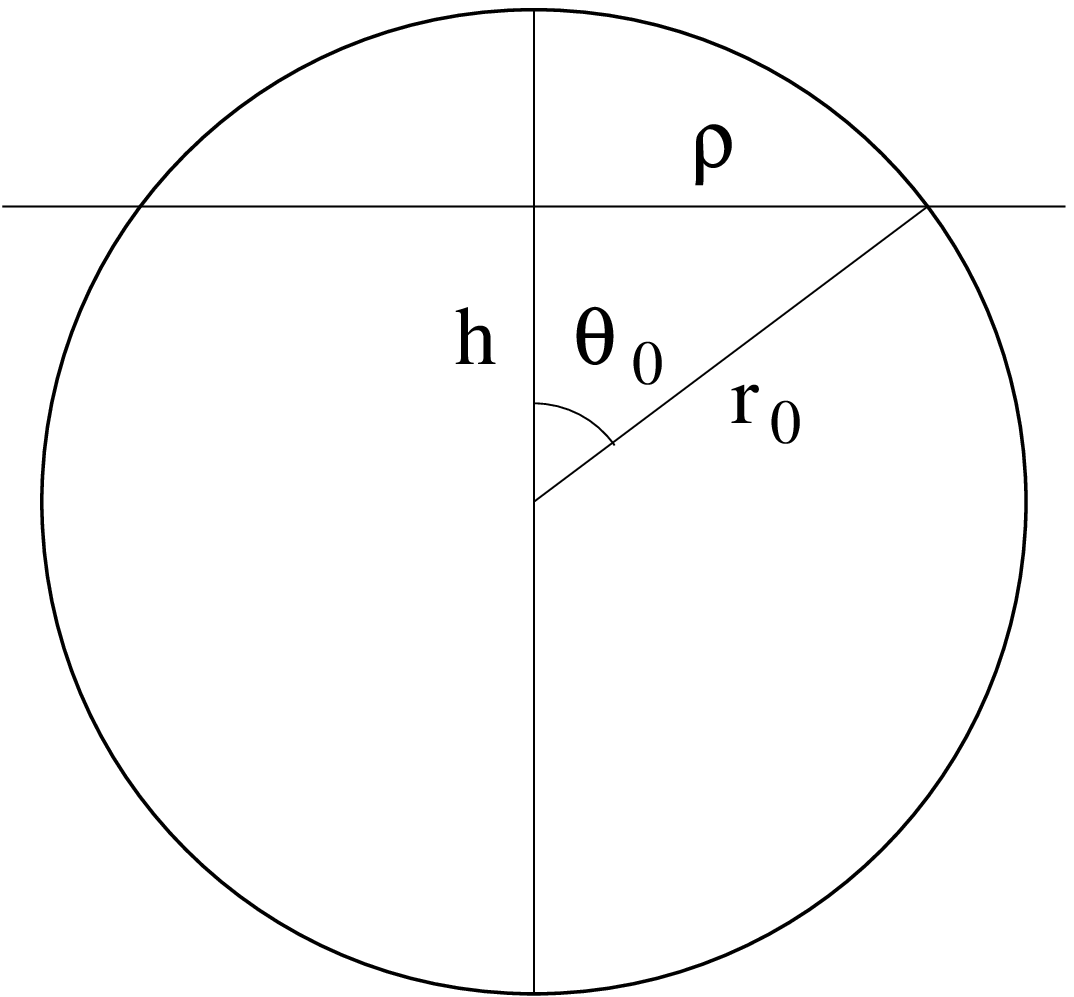,scale=0.4}\\
{\bf Figure 1:} Sphere of radius $r_0$ and water-line for the de\-fi\-ni\-tion
of $h$, $\rho$, and $\theta_0$.}
\smallskip

In section 7 the equations in second order in the deformation are derived. One
finds that only spherical harmonics for the deformation are allowed whose square
projected onto the
space of harmonics with the same $p$ is proportional to the original
deformation. This
is the contents of the shape equations (\ref{shp}). This projection vanishes for
odd $p$. Thus one has to distinguish between the problem for odd and even $p$.
In the following only the case of even $p$, that is for central symmetric
bodies, is pursued. A large number of solutions for the shape equations is found in section 8 assuming
invariance under various subgroups of the orthogonal group O(3). By considering
only these special groups only solutions with mirror symmetry are found. It is
not clear whether there are shapes with inversion symmetry, but without mirror
symmetry.

In section 9 it is shown how contributions in higher order can be obtained,
which lead to a formal expansion for the shape.

\section{General Considerations}

\paragraph{Potential energy} Denoting the
volume of the body by $V$, the volume {\bf a}bove the water by $V_{\rm a}$, that
{\bf b}elow the water by $V_{\rm b}$, one obtains (Archimedes' law)
\be
V_{\rm a} = (1-\dens) V, \quad V_{\rm b} = \dens V,
\ee
where $\dens$ is the relative density of the body with respect to the liquid.
Denote the total mass of the body by $m$,
the masses above/below the water-line by $m_{\rm a,b}$,
the center of mass above/below the water-line by $C_{\rm a,b}$,
and the distance of $C_{\rm a,b}$ from the water-line by $d_{\rm a,b}$.
Then the potential energy ${\cal V}$ of the system is given by
\be
{\cal V} = m_{\rm a} g d_{\rm a} + (m-m_{\rm b}) g d_{\rm b}
=m(1-\dens) g (d_{\rm a}+d_{\rm b}).
\ee
Thus the difference in height between the two centers of mass,
$d=d_{\rm a}+d_{\rm b}$, is constant, since it has to be independent of the
orientation.
This does not imply, that $d_{\rm a}$ and $d_{\rm b}$ are separately constant.
Moreover the line $C_{\rm a} C_{\rm b}$ connecting the two centers of mass has
to be
perpendicular to the water-level.
Placing the center of mass of the body in the origin and denoting the
coordinates of $C_{\rm a,b}$ by $(0,0,z_{\rm a,b})$ one obtains
\be
z_{\rm a}=\dens d, \quad z_{\rm b}=-(1-\dens) d.
\ee
Thus the loci of the centers of gravity lie on spheres.

The reverse is also true: If $z_{\rm a}$ is independent of the orientation, then
also $z_{\rm b}$
has this property. Then $z_{\rm a}-z_{\rm b}=d_{\rm a}+d_{\rm b}$ is constant
and thus its potential
energy. We will use this property to determine bodies which can float in all
orientations.

One can conclude: If a body has the property that a plane in arbitrary
orientation cutting through it, so that the two volumina $V_{\rm a,b}$ are
constant
and the centers of gravity of these volumina lie on spheres, then it is a floating body of equilibrium, since the body assumes in all orientations the same potential energy.

These properties and another one we will derive below were derived by Pierre
Bouguer and Charles Dupin a long time ago. Auerbach writes in his 1938
paper\cite{Auerbach}:
{\it Il r\'esulte ais\'ement des th\'eor\`emes classiques de Bouguer et Dupin
que la
condition n\'ecessaire et suffisante pour qu'un corps soit une solution de ce
probl\`eme est que la surface de centres de car\`ene soit une sph\`ere.
De plus, on peut affirmer que dans ce cas l'ellipse centrale d'inertie de la
flottaisson est un cercle dont le rayon est le m\^eme pour toute position
d'\'equilibre.} Indeed Pierre Bouguer (1698-1758) and Charles Dupin (1784-1873)
wrote books\cite{Bouguer,Dupin} on the hydrostatics of ships. These conditions are also found in the textbooks by Appell\cite{Appell} and Webster\cite{Webster}. Compare also the article by Gilbert\cite{Gilbert}.

\paragraph{Moment of inertia} Rotate the body by an infinitesimal angle. Since the volume above and below the
water has to be conserved the rotation is around the center of gravity M of the
water-plane area $F$ (intersection of the plane of the water-surface with the body). The center of gravity M of the water-plane area is given by
\be
x_{\rm M} = \frac 1F \int\de F\, x, \quad y_{\rm M} = \frac 1F \int\de F\, y.
\ee
An infinitesimal rotation $\delta\phi_{x,y}$ will bring a wedge of thickness
\be
\zeta(x,y)=\delta\phi_x (x-x_{\rm M})+\delta\phi_y (y-y_{\rm M})
\ee
above the waterline, if positive; if negative, its modulus describes the
thickness of a wedge disappearing below the waterline. There are two
contributions to the shift of the centers of gravity in $x$ and $y$ direction,
\bea
\delta x_{\rm a} = -z_{\rm a} \delta\phi_x
+\frac 1{V_{\rm a}} \int \de F\, x \zeta(x,y), &&
\delta y_{\rm a} = -z_{\rm a} \delta\phi_y 
+\frac 1{V_{\rm a}} \int \de F\, y \zeta(x,y), \\
\delta x_{\rm b} = -z_{\rm b} \delta\phi_x
-\frac 1{V_{\rm b}} \int \de F\, x \zeta(x,y), &&
\delta y_{\rm b} = -z_{\rm b} \delta\phi_y
-\frac 1{V_{\rm }b} \int \de F\, y \zeta(x,y),
\eea
where the first term is due to the rotation of the centers of gravity, and the
second one comes from the appearance and disappearance of the wedges. The
requirement
$\delta x_{\rm a}=\delta x_{\rm b}$, $\delta y_{\rm a}=\delta y_{\rm b}$ yields
for arbitrary $\delta\phi_{x,y}$ the conditions
\be
I_{ij} = \delta_{ij} I, \quad I = \dens(1-\dens)Vd \label{cond}
\ee
for the moments of inertia of the water-plane area
\bea
I_{xx} &=& \int\de F\, (x-x_{\rm M})^2, \\
I_{xy} &=& \int\de F\, (x-x_{\rm M})(y-y_{\rm M}), \\
I_{yy} &=& \int\de F\, (y-y_{\rm M})^2.
\eea
The moment $I$ is independent of the orientation.

If (\ref{cond}) is not fulfilled, but both eigenvalues of the matrix
\be
\hat I = \left(\begin{array}{cc} I_{xx} & I_{xy} \\ I_{xy} & I_{yy} \end{array}\right)
\ee
are larger than $I$, then the body is in stable equilibrium. If both are less, then the body is in unstable equilibrium, and if one is larger, one less than $I$, then it is in saddle-point equilibrium.
Here we are interested in an
indifferent equilibrium, for which equality (\ref{cond}) holds.
The corresponding condition yields in two dimensions that the length of the
waterline has to be constant.\cite{WegnerI}

\section{Central Symmetric Case $\dens=1/2$\label{rhohalf}}

In this section central symmetric bodies with relative density $\dens=1/2$ are
considered.
Central symmetry, which is also called inversion symmetry, means: If the point
at $\w r$ belongs to the body, then also the point $-\w r$ belongs to it, where
the center has been placed at the origin. The plane of
the water-plane area goes through the origin in all orientations, since it cuts the body into two equal halves.

The following theorem due to Schneider\cite{Schneiderd,Schneidere} and to
Falconer\cite{Falconer}, also referred to by Hensley in the Scottish
book\cite{Scottish} holds: {\it For arbitrary dimension $d$ and
density $1/2$, if the body is star-shaped, symmetric, bounded and measurable,
then it differs from a ball by a set of measure 0.} It follows from theorem
1.4 of Schneider \cite{Schneidere} (similarly corollary 3.1 in
\cite{Schneiderd}):
'Let $\Omega_d$ be the unit sphere $|\w u|=1$, and $\cdot$ the inner product. If
$\Phi$ is an even real-valued, countably additive set function on $\Omega_d$
satisfying $\int_{\Omega_d} |\w
u\cdot\w v| \de\Phi(u) = 0$ for each $\w v\in\Omega_d$, then $\Phi=0$.'
The theorem will be reproved in the following assuming that the distance of the
surface from the center of the body is a continuous function of $\w u$.
This is done first for $d=3$ dimensions and then generalized to
dimensions $d>3$. The premise that the body is star-shaped is
important. It is shown in subsection (\ref{holes}) that there are
non-spherical floating bodies of equilibrium, if one allows holes to be drilled into the body.

\subsection{Star-shaped body in three dimensions}

If the body is star-shaped, i.e. there exists a point A such that for each
point P in the body the segment AP lies in the body. Since the set of these
points A (called kernel) forms a convex set and since it is central symmetric,
too, the origin is such a point A. We denote the extension of the body from the
origin in direction of the unit vector $\w u$ by $r(\w u)$. Let $\Omega_u$ be
the unit sphere and $\w v$ the normal on the water surface, then
\be
\frac 12 Vd = \int\de^3 r \de\Omega_u r |\w u\cdot \w v|
=\frac 14\int\de\Omega_u r^4(\w u) |\w u\cdot \w v|
\ee
has to be independent of $\w v$. One expands $|\w u\cdot \w v|$ in Legendre
polynomials
\be
|\w u\cdot \w v| = \sum_{n=0} c_n P_{2n}(\w u\cdot \w v), \quad
c_n = (-)^{n+1} \frac{(2n+1/2)\Gamma(n-1/2)}{2\sqrt{\pi}(n+1)!}.
\ee
Note that all coefficients $c_n$ differ from 0, $c_n\not=0$. Using the addition
theorem
\be
P_{2n}(\w u\cdot\w v) = \frac{4\pi}{2n+1} \sum_m Y_{2n,m}(\w u)\,Y_{2n,m}(\w v),
\ee
where the $Y_{2n,m}$ are $4n+1$ orthonormalized (real) spherical harmonics, one obtains
\be
Vd = \sum_{n,m} \frac{2\pi c_n}{2n+1} Y_{2n,m}(\w v)
\int\de\Omega_u r^4(\w u) Y_{2n,m}(\w u).
\ee
$Vd$ is only independent of $\w v$, if all coefficients $c_n$ except $c_0$
vanish.
Since the functions $Y_{2n,m}$ form a complete orthogonal set of central
symmetric functions, this implies that $r^4(\w u)$ has to be independent of
$\w u$.

\subsection{Star-shaped body in higher dimensions}

The proof is similar to that in $d$ dimensions. $r^3$ and $r^4/4$ have to be
replaced by $r^d$ and $r^{d+1}/(d+1)$, resp.
Instead of Legendre polynomials one has to expand in ultraspherical (Gegenbauer)
polynomials $G^{(d/2-1)}_n(\w u\cdot \w v)$,
\bea
|\w u\cdot \w v| &=& \sum_{n=0}^{\infty} c_n G_{2n}^{(d/2-1)}(\w u\cdot \w v), \nn
c_n &=& \frac{(-)^{n+1}(2n+d/2-1)\Gamma(n-1/2)\Gamma(d/2-1)}
{2\pi\Gamma(n+d/2+1/2)}.
\label{expz}
\eea
Repeated application of Gegenbauer's addition theorem (see appendix \ref{Gegenbauer}) yields
\be
G_n^{(d/2-1)}(\w u \cdot \w v) = \frac{2\pi^{d/2}}{(n+d/2-1)\Gamma(d/2-1)}
\sum_m Y^{(d)}_{n,m}(\w u) Y^{(d)}_{n,m}(\w v), \label{addY}
\ee
where the functions $Y^{(d)}_{n,m}$ form a complete set of orthonormal functions on
$\Omega_d$
obeying
\be
\triangle_u Y^{(d)}_{n,m}(\w u) = -n(n+d-2) Y^{(d)}_{n,m}(\w u), \label{Lapld}
\ee
with the Laplacean operator $\triangle_u$ on the sphere $\Omega_d$.
Since all $c_n\not=0$ the same argument applies that $r^{d+1}(\w u)$ has to be
independent of $\w u$.

\subsection{Bodies with holes\label{holes}}

If the requirement that the body is star-shaped is omitted, but holes in
the body are allowed, then the condition that the body K
can float in all orientations requires that the integral
$\int_0^{\infty} \de r \chi(r,\w u) r^d$
does not depend on the direction $\w u$, where the indicator function
\be
\chi(r,\w u) = \left\{ \begin{array}{cl}
1 & r\w u \in\,{\rm K} \\
0 & r\w u \not\in\,{\rm K}.
\end{array} \right.
\ee
is introduced. Thus one may drill holes into the body and add the corresponding
amount of material at the outside creating a bulge.

\begin{figure}[ht]
\parbox[t]{6cm}
{\epsfig{file=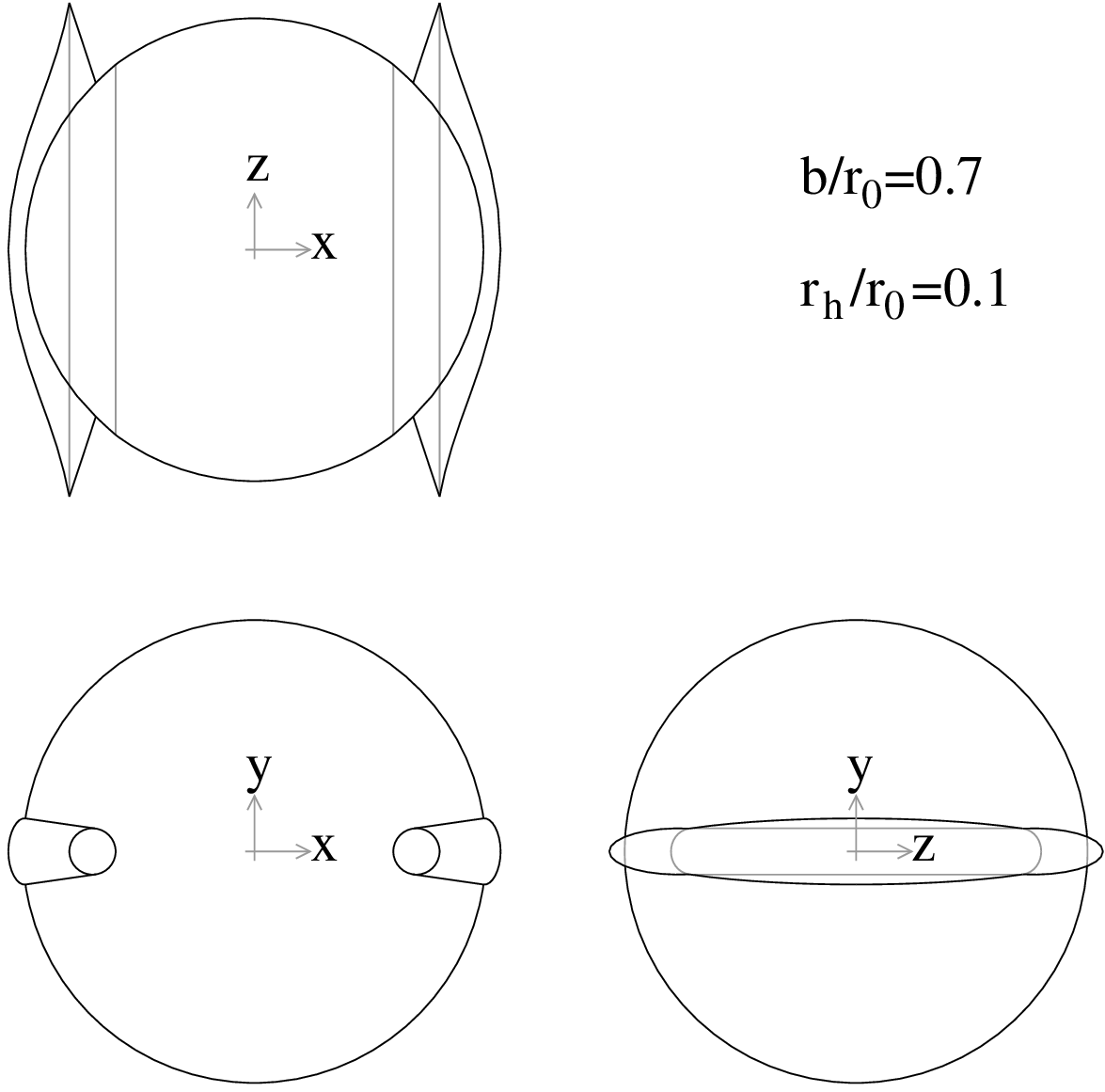,angle=180,scale=0.5}}
\hfill
\parbox[t]{5cm}{{\bf Figure 2:} Non-spherical central symmetric floating body
of equilibrium with two holes described in this
section at $\dens=1/2$ projected onto the (x,y)-, (x,z)-, and (y,z)-plane.\\
The inner part of the cylindrical holes end at the sphere, the outer part at the
bulge.}
\end{figure}

As an example consider two circular cylindrical holes of radius $r_h$ around the
axes $x=\pm b$, $y=0$ drilled into a sphere of radius $R_0$. Using spherical
coordinates
\be
x=r\sin\theta\cos\phi, \quad y=r\sin\theta\sin\phi, \quad z=r\cos\theta
\ee
the bounds of the hole is given by
\be
r=R_{\pm}=\frac{b\cos\phi \pm \sqrt{r_h^2-b^2\sin^2\phi}}{\sin\theta}.
\ee
One obtains the indicator function
\newcommand{\ttheta}{\tilde\theta}
\be
\chi(r,\theta,\phi) = \left\{\begin{array}{ll}
\ttheta(R_--r) + \ttheta(R'-r)\ttheta(r-R_+)
& {\rm if}\,\, |\sin\phi|<\frac{r_h}d \,\, {\rm and}\,\, R_-<R_0, \\
\ttheta(R_0-r) & {\rm otherwise}
\end{array} \right.
\ee
with the Heaviside step function $\ttheta$, which equals 1 for positive argument
and 0 for negative argument. $\ttheta(0)$ may be chosen 1 (0) for a closed (an
open) body. $R'$ is given by
\be
R' = \sqrt[4]{R_0^4+R_+^4-R_-^4}.
\ee
Obviously there are bodies with several holes and the holes need not be
circular cylinders.
Thus there is a large variety of bodies with central symmetry at density
$\dens=1/2$, if one does not require star-shape.
The body described here has the property that any fluid above the center can
flow off and any air below the center can climb up up. In two dimensions a body
with this property is not possible, since two holes through the body would cut
it into three pieces. Otherwise, however, fluid above the center cannot always outflow nor air below the center cannot always bubble up.

\section{Parametrization of the Surface}

In the following bodies of general densities are considered, however,
star-shape is assumed.
Thus in spherical coordinates the radius $r$ of the surface of the body
is a unique function of the angles $\theta_{\rm K}$, $\phi_{\rm K}$ in the body-fixed system.
 
It is expanded in the set of spherical harmonics
$C_{l,m}(\theta_{\rm K}) \ex{\ie m \phi_{\rm K}}$ (Racah uses this normalization
for $C$, however here $C$ contains only the $\theta$-dependence),
\be
r = \sum_{l,m} a_{l,m} C_{l,m}(\theta_{\rm K}) \ex{\ie m \phi_{\rm K}},
\ee
which are related to the conventional spherical harmonics by
\be
Y_{l,m}(\theta_{\rm K},\phi_{\rm K})
= \sqrt{\frac{2l+1}{4\pi}} C_{l,m}(\theta_{\rm K}) \ex{\ie m \phi_{\rm K}}.
\ee
In contrast to section \ref{rhohalf} and appendix \ref{Gegenbauer}, where the
spherical harmonics were assumed to be real, here the spherical harmonics
conventionally applied in physics are used\cite{Wignerd,Wignere,Edmonds}.
In particular one has
\be
C_{00} = 1, \quad C_{10}(\theta) = \cos(\theta), \quad
C_{1,\pm 1}(\theta) = \mp \frac 1{\sqrt 2} \sin\theta.
\ee
and in general
\be
C_{lm}(\theta) = (-)^m \sqrt{\frac{(l-m)!}{(l+m)!}} P_{lm}(\cos\theta)
\ee
holds with the Legendre functions defined by
\be
P_{lm}(x) = (1-x^2)^{m/2}\frac 1{2^ll!} \frac{\de^{l+m}}{\de x^{l+m}} (x^2-1)^l.
\ee
The following relation holds
\be
C_{l-m}(\theta) = (-)^m C_{lm}(\theta).
\ee

The body is rotated by means of Euler angles $(\alpha,\beta)$ into a position
where the water-surface is parallel to the $x,y$-plane. A third rotation around
the axis perpendicular to the water-surface is not necessary. The functions
$C_{lm'}(\theta_{\rm K})\ex{\ie m' \phi_{\rm K}}$ transform according to
\be
C_{lm'}(\theta_{\rm K})\ex{\ie m' \phi_{\rm K}}
= \sum_m D^l_{mm'}(\alpha,\beta,0) C_{lm}(\theta)\ex{\ie m\phi},
\ee
from which we obtain the radius
\be
r(\theta,\phi;\beta,\alpha) = \sum_{l,m,m'} C_{lm}(\theta)
\Dphi a_{l,m'}
\ee
with the Wigner rotation matrix elements\cite{Edmonds,Wignerd,Wignere}
\be
\Dphi
= \ex{\ie m \phi} d^l_{mm'}(\beta) \ex{\ie m'\alpha}.
\ee

The height of the water level above the center of mass of the body be
$h(\beta,\alpha)$. Then the intersection of the water level with the surface of
the body is given by
\be
h(\beta,\alpha) = \cos(\Theta) r(\Theta,\phi;\beta,\alpha).
\label{eqh}
\ee
This determines $\Theta(\phi,\beta,\alpha)$ of the waterline on the
surface of the body for given
orientation $(\alpha,\beta)$. At water-level the waterline is given in 
polar coordinates $(\rho,\phi)$ by
\be
\rho(\phi,\beta,\alpha)
= \sin(\Theta) r(\Theta,\phi;\beta,\alpha).
\label{eqrho}
\ee
The total volume $V$ and the volume $V_{\rm a}$ above the water surface are
given by
\bea
V &=& \frac 13 \int\de\phi\int_0^{\pi} \de\theta r^3\sin\theta, \label{Vint} \\
V_{\rm a} &=& 
\frac 13 \int\de\phi\int_0^{\Theta(\phi)} \de\theta r^3\sin\theta
-\frac h6 \int\de\phi\rho^2. \label{Vaint}
\eea
The volume $V_{\rm a}$, which is a segment, is divided
into a sector (first integral) and a cone (second integral). 
Obviously one has $V_{\rm b}=V-V_{\rm a}$.
Similarly one obtains the coordinates $z$ of the centers of gravity by means of
$Z_{\rm c}=Vz_{\rm c}$ and $Z_{\rm a,b}=V_{\rm a,b} z_{\rm a,b}$ from
\bea
Z_{\rm c} &=& \frac 14 \int\de\phi\int_0^{\pi}\de\theta r^4\sin\theta\cos\theta,
\label{Zcint} \\
Z_{\rm a} &=& \frac 14 \int\de\phi\int_0^{\Theta(\phi)}\de\theta
r^4\sin\theta\cos\theta
-\frac{h^2}8 \int\de\phi \rho^2. \label{Zaint}
\eea
The integrals (\ref{Vint}-\ref{Zaint}) can often be expressed in the form
\bea
\int\de\phi\sum_{lmm'} \Dphi \hat f_{lm'}
&=& 2\pi \sum_{lm'} \Dzeros \hat f_{lm'} \nn
&=& 2\pi \sum_{lm'} C_{lm'}(\beta) \ex{\ie m'\alpha} \hat f_{lm'}.
\eea
Since the integrals must not depend on of $\alpha,\beta$ all $\hat f_{lm'}$ have
to vanish with the exception of $\hat f_{00}$.

\subsection{The Sphere}

For a sphere of radius $r_0$ with the waterline at
$\Theta=\theta_0$ one obtains (see figure 1)
\be
h=r_0\cos\theta_0, \quad \rho=r_0\sin\theta_0,
\ee
independent of the orientation. Since the cosine and
sine of $\theta_0$ is often needed, the abbreviations
\be
\ct=\cos\theta_0, \quad \st=\sin\theta_0
\ee
are used. The volumes are
\bea
V_{\rm a} &=& \frac{\pi}3 r_0^3 (2-3\ct+\ct^3)
=\frac{\pi}3 r_0^3 (1-\ct)^2(2+\ct) \nn
&=& \frac{\pi}3 (2r_0^3 -3r_0^2h+h^3), \label{va0} \\
V_{\rm b} &=& \frac{\pi}3 r_0^3 (2+3\ct-\ct^3)=
\frac{\pi}3 r_0^3 (1+\ct)^2 (2-\ct) \nn
&=& \frac{\pi}3 (2r_0^3 +3r_0^2h-h^3),
\eea
and the $z$-coordinates
\be
z_{\rm a} =\frac{3r_0}4 \frac{(1+\ct)^2}{2+\ct}, \quad
z_{\rm b} =-\frac{3r_0}4 \frac{(1-\ct)^2}{2-\ct} \label{za0}
\ee
One easily calculates
\be
d=z_{\rm a}-z_{\rm b} = \frac{3r_0}{(2+\ct)(2-\ct)}, \quad I=\frac{\pi}4 \rho^4
\ee
and verifies eq. (\ref{cond}).

\section{Expansion}

Starting out from the sphere the body will be deformed and an expansion in
the deformation parameter $\epsilon$ will be performed. That is, $a$ and
similarly the other quantities are expanded in powers of $\epsilon$,
\be
a_{l,m} = \sum_{n} \epsilon^n a_{n;l,m}.
\ee
Starting point of the expansion is the sphere of radius $r_0$,
\be
a_{0;l,m} = r_0 \delta_{l,0} \delta_{m,0}.
\ee
Similarly $r$ is expanded,
\be
r(\theta,\phi;\beta,\alpha) = \sum_n \epsilon^n r_n(\theta,\phi;\beta,\alpha)
\ee
with
\be
r_n(\theta,\phi;\beta,\alpha) 
= \sum_{lmm'} C_{lm}(\theta) \Dphi a_{n;lm'}
\ee
Since $r_n$ has to be evaluated around $\theta=\theta_0$, the
expansion
\be
r_n(\theta,\phi;\beta,\alpha) = \sum_k (\theta-\theta_0)^k
r_{nk}(\phi,\beta,\alpha) \label{thetaexp1}
\ee
with
\be
r_{nk}(\phi,\beta,\alpha)
= \frac 1{k!} \sum_{lmm'} \left. \frac{\de^k C_{lm}(\theta)}{\de\theta^k}
\right|_{\theta=\theta_0} \Dphi a_{n;lm'} \label{thetaexp2}
\ee
is used.
$r_0$ does not depend on $\theta,\phi,\beta,\alpha$. Similarly
$\Theta(\phi)$, $h$ and $\rho$ are expanded in powers of $\epsilon$,
\bea
\Theta(\phi;\beta,\alpha)
&=& \sum_n \epsilon^n \Theta_n(\phi;\beta,\alpha), \\
h(\beta,\alpha) &=& \sum_n \epsilon^n h_n(\beta,\alpha), \\
\rho(\phi,\beta,\alpha) &=& \sum_n \epsilon^n \rho_n(\phi,\beta,\alpha).
\eea
$\Theta_0=\theta_0$, $h_0=r_0\ct$, $\rho_0=r_0\st$ do not depend on the
angles, since the body is a sphere in lowest (zeroth) order.

\subsection{Evaluation of the Integrals \label{evint}}

In order to evaluate the integrals for the part of the body above the
water-level I divide the integral over $\theta$ into one which runs up to
$\theta_0$ and the one from $\theta=\theta_0$ to $\Theta(\phi)$.
Let us first consider the integral
up to $\theta_0$. For the volume we expand in powers of $\epsilon$ and in
spherical harmonics
\be
r^3/3=\sum_n \epsilon^n f_n
= \sum_{nlm} \epsilon^n C_{lm}(\theta) \ex{\ie m \phi} f_{n;lm}.
\ee
Due to the $\phi$-integration only terms with $m=0$ will contribute.
With $\zeta=\cos\theta$ one obtains
\bea
\int_0^{\theta_0} \de \theta C_{l0}(\theta) \sin\theta
=\int_{\ct}^1 \de \zeta\, P_l(\zeta) \\
=\left. \sqrt{1-\zeta^2} P_{l,-1}(\zeta) \right|_{\cos\theta}^1
=-\st P_{l,-1}(\ct).
\eea
Thus the first contribution $V_{{\rm a}n}^{(1)}$ reads
\be
V_{{\rm a}n}^{(1)}=\int\de\phi\int_0^{\theta_0} \sin\theta f_n
=2\pi\left( (1-\ct) f_{n;00}
-\st \sum_{l\ge 1} f_{n;l0} P_{l,-1}(\ct)\right). \label{V1}
\ee
The term for $l=0$ is written separately, since $P_{0,-1}$ is not
defined.
The total volume yields
\be
V_n = 4\pi f_{n;00}.
\ee
Next the corresponding integral for $Z$ is considered. The expansion
\be
r^4/4 = \sum_n \epsilon^n f^z_n
= \sum_{nlm} \epsilon^n C_{lm}(\theta) \ex{\ie m \phi} f^z_{n;lm}.
\ee
yields
\bea
\int_0^{\theta_0} \de \theta C_{l0}(\theta) \sin\theta \cos\theta
&=& \int_{\ct}^1 \de \zeta\, \zeta P_l(\zeta) \\
&=&\left. \zeta \sqrt{1-\zeta^2} P_{l,-1}(\zeta) \right|_{\ct}^1
-\int_{\ct}^1 \sqrt{1-\zeta^2} P_{l,-1}(\zeta) \nn
&=&\left. (\zeta \sqrt{1-\zeta^2} P_{l,-1}(\zeta)
-(1-\zeta^2) P_{l,-2}(\zeta) ) \right|_{\ct}^1 \\
&=&-\st\ct P_{l,-1}(\ct)
+\st^2 P_{l,-2}(\ct).
\eea
Thus the first contribution to $Z^{(1)}_{{\rm a}n}$ reads
\bea
Z^{(1)}_{{\rm a}n} &=& \int\de\phi\int_0^{\theta_0}\de\theta f^z_n \sin\theta
\cos\theta
= 2\pi\left( \frac 12 (1-\ct^2) f^z_{n;00} + \frac 13 (1-\ct^3)
f^z_{n;10} \right. \nn
&+& \left. \st \sum_{l\ge 2} f^z_{n;l0} (\st P_{l,-2}(\ct) -\ct
P_{l,-1}(\ct))\right). \label{Z1}
\eea
For the complete volume one obtains
\be
Z_{{\rm c}n} = \frac{4\pi}3 f^z_{n;10}.
\ee
We require that the origin coincides with the center of mass of the complete
body. This yields $f^z_{n;1,0}=0$. This can always be done by an appropriate
choice of the $a_{n;1,m}$.

Since $\Theta(\phi)-\theta_0=O(\epsilon)$ the integral over the interval
$\theta=\theta_0 ... \Theta(\phi)$ can be expanded in powers of
$\epsilon$.
Thus the second contributions to $V_{\rm a}$ and $Z_{\rm a}$ yield
\be
V_{{\rm a}n}^{(2)} = \int\de\phi g_n \label{V2}
\ee
with
\be
\sum_n \epsilon^n g_n
= \frac 13 \int_{\theta_0}^{\Theta(\phi)} \de\theta r^3\sin\theta
-\frac h6 \rho^2.
\ee
Similarly the second contribution to $Z_{\rm a}$ is obtained from
\be
Z_{{\rm a}n}^{(2)} = \int\de\phi g^z_n \label{Z2}
\ee
with
\be
\sum_n \epsilon^n g^z_n
=\frac 14 \int_{\theta_0}^{\Theta(\phi)} \de\theta r^4 \sin\theta
\cos\theta
-\frac 18 h^2\rho^2.
\ee
It is useful to perform the expansion procedure
explicitly in zeroth, first and second order. Then it becomes apparent how
to continue to higher orders.

\subsection{Zeroth Order: The Sphere again}

In zeroth order one obtains
\bea
f_0 = \frac 13 r_0^3, && f^z_0 = \frac 14 r_0^4, \\
g_0 = -\frac 16 r_0^3 \ct\st^2, && g^z_0 = -\frac 18 r_0^4 \ct^2 \st^2.
\eea
One obtains from these expressions
\bea
V_{{\rm a}0}^{(1)} &=& 2\pi (1-\ct) f_{0;00} = \frac{2\pi}3 r_0^3(1-\ct), \\
V_{{\rm a}0}^{(2)} &=& -\frac{\pi}3 r_0^3 \ct \st^2,
\eea
which yields the volume in agreement with (\ref{va0}), and
\bea
Z_{{\rm a}0}^{(1)} &=& \pi (1-\ct^2) f^z_{0;00} = \frac{\pi}4 (1-\ct^2) r_0^4,
\\
Z_{{\rm a}0}^{(2)} &=& -\frac{\pi}4 (1-\ct^2)\ct^2 r_0^4,
\eea
which yields $z_{\rm a}$ in eq. (\ref{za0}).

\section{First order in the deformation}

The deformation of the sphere in first order in $\epsilon$ is considered in this section.
One obtains from eqs. (\ref{eqh},\ref{eqrho})
\be
\Theta_1 = \frac{r_{10} \ct-h_1}{r_0\st}, \quad
\rho_1 = \frac{r_{10}-h_1\ct}{\st}.
\ee
Following the procedure in subsection \ref{evint} one obtains
$f_1 =r_0^2r_1$ or more explicitly
\be
f_1=\sum_{lm} f_{1;lm} C_{lm}(\theta) \ex{\ie m\phi}, \quad
f_{1;lm}=r_0^2 \sum_{m'} D^{(l)}_{m,m'}(\alpha,\beta,0) a_{1;lm'}
\ee
and thus the volume
\be
V_1 = 4\pi r_{1;00} r_0^2.
\ee
We choose to keep the volume unchanged in this order, which implies
\be
r_{1;00}=a_{1;00}=0.
\ee
Next one obtains $g_1 = -\frac 12 h_1r_0^2 \st^2$ and thus
\be
V_{{\rm a}1} = -2\pi \st r_0^2 
\sum_{l\ge 1} r_{1;l0} P_{l,-1}(\ct) -\pi h_1r_0^2 \st^2.
\ee
Since $V_{{\rm a}1}$ is constant, one obtains
\be
h_1=h_{1;00}-\frac 2{\st} \sum_{l\ge 1} r_{1;l0} P_{l,-1}(\ct) \label{h1}
\ee
with the $\alpha,\beta$ independent contribution $h_{1;00}$.

Next one determines $Z$. With $f^z_1 = r_0^3r_1$ one obtains
\be
Z_{{\rm c}1}=\frac{4\pi}3 r_0^3 r_{1;10}.
\ee
Since the center of gravity stays at the origin, one requires $r_{1;10}=0$, which is equivalent to $a_{1;1m}=0$.
With $g^z_1 = -\frac 12 h_1 r_0^3 \ct\st^2$ one obtains
\be
Z_{{\rm a}1} = \pi \st r_0^3 \left( 2\sum_{l\ge 2} r_{1;l0} (\st P_{l,-2}(\ct)
-\ct P_{l,-1}(\ct)) -h_1 \ct\st \right).
\ee
Inserting eq. (\ref{h1}) into the expression for $Z_{{\rm a}1}$ yields
\be
Z_{{\rm a}1} = \pi \st^2 r_0^3 \left(2\sum_{l\ge 2} r_{1;l0} P_{l,-2}(\ct)
-h_{1;00} \ct \right)
\ee
$Z_{{\rm a}1}$ must be constant. We note that
\be
r_{1;l0} = \sum_{m} \Dzero a_{1;lm}.
\ee
In order that $Z_{{\rm a}1}$ is constant all coefficients
$r_{1;l0} P_{l,-2}(\ct)$
except for $l=0$ have to vanish. Therefore for any given $l$ either all
coefficients $a_{1;lm}$ or $P_{l,-2}(\ct)$ have to vanish. We note that
$P_{l,-2}(\pm 1)=0$ for all $l$. However this corresponds to the limits
$\dens=0$ and $\dens=1$, in which case only a sphere is possible\cite{Montejano}. Further one has $P_{l,-2}(0)=0$ for all odd $l$. This
corresponds to
$\dens=1/2$. It may well be that similarly to the two-dimensional case one has a large variety of solutions for this special density. This case will not be
considered further in this paper.

In the following only cases are considered, for which $\ct$ is different from 0 and $\pm 1$.
We will assume a $\cos\theta_0=\ct$ for which $P_{p,-2}(\ct)=0$ and assume that
for this $\ct$ there is only one solution $p$.
(Numerical calculation up to $l=100$ shows that apart from
$\ct=0,\pm 1$ there is no $\ct$, which for two different $l$s yields $P_{l,-2}(\ct)=0$.)
Thus all $a_{1;lm}=0$ with the exception of $l=p$ and one obtains
\be
r_1=\sum_m C_{pm}(\theta) D^p_{mm'}(\alpha,\beta,\phi) a_{1;pm'},
\ee
where the amplitudes $a_{1;pm'}$ are as yet undetermined. Similarly $h_{1;00}$ is undetermined. Note that the zeroes of $P_{p,2}(x)$ and $P_{p,-2}$ are identical.

\section{Second Order in the deformation}

One obtains from eqs. (\ref{eqh}, \ref{eqrho}) in second order in $\epsilon$
\bea
\Theta_2 &=& \frac 1{2r_0^2\st^3}
(-2\st^2 h_2r_0 +2\ct\st^2 r_{20} r_0 -2\ct\st r_{11} h_1 \nn
&&+2\ct^2\st r_{11} r_{10}- \ct h_1^2 -\ct(2-\ct^2) r_{10}^2 +2h_1r_{10}),
\\
\rho_2 &=& \frac 1{2r_0\st^3}
(-2\ct\st^2 h_2r_0 +2\st^2 r_{20}r_0 -2\st r_{11} h_1 \nn
&&+2\st\ct r_{11} r_{10} -h_1^2 -\ct^2 r_{10}^2 +2\ct h_1r_{10}).
\eea
These quantities depend on $\phi$, $\alpha$ and $\beta$. Quantities on the right
hand side depending on $\theta$ have to be evaluated at $\theta=\theta_0$. The
deviation of $\theta$ from $\theta_0$ is taken into account by the derivatives
$r_{11}$ with respect to $\theta$, compare eqs. (\ref{thetaexp1}, \ref{thetaexp2}).
For products $b^{(1)} b^{(2)}$ of
\be
b^{(i)} = \sum_{lm} b^{(i)}_{lm} C_{lm}(\theta) \ex{\ie m \phi}
\ee
we use the notation
\be
b^{(1)} b^{(2)}
= \sum_{lm} (b^{(1)} b^{(2)})_{;lm} C_{lm}(\theta) \ex{\ie m \phi}
\ee
with
\be
(b^{(1)} b^{(2)})_{;lm} = \sum_{l_1,m_1,l_2,m_2} \langle l0;l_10,l_20 \rangle
\langle lm;l_1m_1,l_2m_2 \rangle b^{(1)}_{l_1m_1} b^{(2)}_{l_2m_2},
\ee
where the $\langle lm;l_1m_1,l_2m_2 \rangle$ are Clebsch-Gordan coefficients.
Use is made of
\be
C_{l_1,m_1}(\theta) C_{l_2,m_2}(\theta) = \sum_{lm}
\langle lm;l_1m_1,l_2m_2 \rangle \langle l0;l_10,l_20 \rangle C_{l,m}(\theta).
\ee
Then one obtains from
\bea
f_2 &=& r_0r_1^2+r_0^2r_2, \\
f^z_2 &=& r_0^3 r_2 + \frac 32 r_0^2 r_1^2
\eea
the expressions
\bea
f_{2;lm} &=& r_0 \sum_{m'} D^{(l)}_{m,m'}(\alpha,\beta,0)
( (a_1^2)_{;lm'} + r_0 a_{2;lm'} ), \\
f^z_{2;lm} &=& r_0^2 \sum_{m'} D^{(l)}_{m,m'}(\alpha,\beta,0) (\frac 32 (a_1^2)_{;lm'} + r_0 a_{2;lm'}). \label{fz2}
\eea
For the complete volume one obtains
\bea
V_2 &=& 4\pi f_{2;00} = 4\pi (r_0^2 a_{2;00} + \frac{r_0}{2p+1} \sum_m (-)^m
a_{1;pm} a_{1;p-m}), \\
Z_{{\rm c}2} &=& \frac{4\pi}3 f^z_{2;1,0}.
\eea
Since the center of mass should stay at the origin and since $(a_1^2)_{;lm'}$
vanishes for odd $l$ one has $(a_1^2)_{;1m}=0$ and requires $a_{2;1m}=0$.

Then the first contributions to $V_{{\rm a}2}$ and $Z_{{\rm a}2}$ yield
\bea
V^{(1)}_{{\rm a}2} &=& 2\pi \left((1-\ct) f_{2;00} -\st \sum_{l\ge 2} f_{2;l0}
P_{l,-1}(\ct)\right), \\
Z^{(1)}_{{\rm a}2} &=& 2\pi \left(\frac 12 (1-\ct^2)f^z_{2;00}
+\frac 13(1-\ct^3) f^z_{2,10} \right. \nn
&& + \left. \st \sum_{l\ge 2} f^z_{2;l0} (\st P_{l,-2}(\ct)-\ct P_{l,-1}(\ct)
)\right)
\eea
with eqs. (\ref{V1}, \ref{Z1}, \ref{fz2}).

The second contributions are obtained from
\bea
g_2 &=& \frac{r_0}2 (-\st^2 h_2r_0 + \ct r_1^2 -2r_1h_1 + \ct h_1^2), \\
g^z_2 &=& \frac 14 r_0^2 (-2\ct\st^2 h_2r_0 +2\ct^2 r_1^2 -4\ct r_1h_1
+(3\ct^2-1)h_1^2).
\eea
together with eq. (\ref{V2} ,\ref{Z2}). The condition that $V_{{\rm a}2}$ is independent
of $\alpha,\beta$ yields an equation for $h_2$. Thus $h_2$ is determined up to
some additive $\alpha,\beta$-independent contribution $h_{2;00}$.
Instead of considering $Z_{{\rm a}2}$ it is more practical to consider the difference $Z_{{\rm a}2}-\ct r_0 V_{{\rm a}2}$
\bea
Z^{(2)}_{{\rm a}2}-\ct r_0 V^{(2)}_{{\rm a}2} &=& {\rm const}
-2\pi r_0^2 P^2_{p,-1}(\ct) \sum_{lm} D^l_{0m} (a_1^2)_{;lm} \nn
&+& 2\pi r_0^2 h_{1;00} \st P_{p,-1}(\ct) \sum_m D^p_{0m} a_{1;pm}.
\eea
Then one obtains
\bea
&& Z_{{\rm a}2}-\ct r_0 V_{{\rm a}2} - {\rm const} \\
&=& 2\pi r_0^2\sum_{l\ge2,m} \Dzero
\left( A_{1l}r_0a_{2,lm} + A_{2l}h_{1;00}a_{1,lm} + A_{3l} (a_1^2)_{;lm} \right) \nonumber
\eea
with
\bea
A_{1l} &=& s_0^2 P_{l,-2}(\ct), \\
A_{2l} &=& s_0 \delta_{p,l} P_{p,-1}(\ct), \\
A_{3l} &=& \frac 32 \st^2 P_{l,-2}(\ct) -\frac 12 \st\ct P_{l,-1}(\ct) -
P^2_{p,-1}(\ct).
\eea
From these expressions $a_{2;lm}$ is determined for $l\not= 0,1,p$
\be
r_0 \st^2 P_{l,-2}(\ct) a_{2;lm} =
(-\frac 32 \st^2 P_{l,-2}(\ct) + \frac 12 \ct\st P_{l,-1}(\ct)
+ P^2_{p,-1}(\ct)) (a_1^2)_{;lm}.
\ee
For $l=p$ one obtains the equation
\be
\st h_{1;00} a_{1;pm} = (\frac 12 \ct\st + P_{p,-1}(\ct))
(a_1^2)_{;pm}.
\ee
For odd $p$ the only non-trivial solution can be obtained for $h_{1;00}=0$, since the right-hand side vanishes for odd $p$. For even $p$ a set of
quadratic equations
\be
\gamma a_{1;pm} = \sum_{m_1,m_2} \langle pm; pm_1,pm_2 \rangle a_{1;pm_1}
a_{1;pm_2} \label{shp}
\ee
is obtained with the constant
\be
\gamma = \frac{\st h_{1;00}}{(\frac 12 \ct\st+P_{p,-1}(\ct))
\langle p0; p0,p0 \rangle}.
\ee
I call the equations (\ref{shp}) shape-equations.
They determine the possible shapes of the body in first order in $\epsilon$.
Descriptively it means that the projection of the square of the deformation $a_1$ onto the harmonics with $l=p$ is proportional to the deformation. I expect that there are cubic shape-equations for odd $p$.

The denominator $\frac 12 \ct\st+P_{p,-1}(\ct)$ can be expressed as
$\st^3\ct K_p(\ct)$ with $K_p$ an even polynomial of order $p-4$ in $\ct$.
A recurrence relation for $K_p$ is derived in appendix \ref{dengamma}. Apparently
$K_p(\ct)$ is positive for real argument $\ct$ and even $p\ge 4$. I do
not have a proof for general $p$. However, numerical calculations show that up to $p=36$ the functions $K_p(\ct)$ do not have any real zero.

The solutions of these equations are invariant under rotations. If $a_{1;pm}$ is a solution, then also
\be
a'_{1;pm} = \sum_{m'} D^p_{mm'}(\alpha,\beta,\gamma) a_{1;pm'}
\ee
is solution for arbitrary Euler angles $\alpha$, $\beta$, $\gamma$.

\renewcommand{\topfraction}{0.95}
\renewcommand{\textfraction}{0.05}

\section{Solution of the Shape Equations}
 
The complete solutions of the shape-equations (\ref{shp}) is not known to me. After all they constitute a set of
$2p+1$ quadratic equations. However, a number of special solutions can be given
by restricting to shapes invariant under subgroups of the rotation
group. Under a subgroup of O(3) a certain set of functions indexed by~$i$
\be
\sum_m v_{im} C_{pm}(\theta_{\rm K})\ex{\ie m\phi_{\rm K}}
\ee
is invariant. Thus the amplitude $a_{1;pm}$ is a linear combination of the
$v_{i,m}$s,
\be
a_{1;pm} = \gamma \sum_i \alpha_i v_{i,m}.
\ee
The $v$s should be orthonormal, $\sum_m v^*_{jm} v_{im} = \delta_{ji}$.
Then the shape equations read
\bea
\sum_i \alpha_i v_{i,m} &=& \sum_{i_1i_2} \alpha_{i_1} \alpha_{i_2} \\
&\times& \sum_{m_1m_2} \sqrt{2p+1} (-)^{m_1+m_2}
\left(\begin{array}{ccc} p & p & p \\ m_1 & m_2 & -m \end{array}\right)
v_{i_1,m_1} v_{i_2,m_2}. \nonumber
\eea
Multiplication by $v^*_{j,m}$ summing over $m$ and using orthonormality yields
\bea
\alpha_j &=& \sum_{j_1j_2} w_{j,i_1,i_2} \alpha_{i_1} \alpha_{i_2},
\label{alphaeq} \\
w_{j,i_1,i_2} &=& \sqrt{2p+1} \\
&\times& \sum_{mm_1m_2}
\left(\begin{array}{ccc} p & p & p \\ m_1 & m_2 & -m \end{array}\right)
(-)^{m_1+m_2} v^*_{j,m} v_{i_1,m_1} v_{i_2,m_2}. \nonumber
\eea
If we require $v^*_{j,m} = (-)^m v_{j,-m}$, which implies real $\alpha$,
then we obtain after replacing $-m$ by $m$ under the sum
\be
w_{j,i_1,i_2} = \sqrt{2p+1} \sum_{mm_1m_2}
\left(\begin{array}{ccc} p & p & p \\ m_1 & m_2 & m \end{array}\right)
v_{j,m} v_{i_1,m_1} v_{i_2,m_2}.
\ee
Since the $3j$ Wigner symbol is invariant under change of the sign of all $m$s
and the product of the $v$s changes into its conjugate complex under this
change of sign, $w$ is real. Moreover it is invariant under permutation of the
indices $j$, $i_1$, $i_2$.

We consider two classes of subgroups of the O(3). The first class contains
shapes with an $n$ fold rotation axis, the second one shapes with tetrahedral,
octahedral and icosahedral symmetry. Obviously the shapes of the last class are
special shapes of the first class of groups. Without loss of generality I put $\gamma=1$ in this section. Otherwise one should read $a/\gamma$ instead of $a$.

\subsection{Solutions with rotational symmetry around the $z$-axis}

\begin{table}[t]
\assigncounter{tabcyclg}
\centerline{
\begin{tabular}{|cc|cccc|c|} \hline
\multicolumn{7}{|c|}{\bf Table \arabic{tabcyclg} Solutions of the shape equations}\\
\multicolumn{7}{|c|}{\bf for the dihedral groups $D_{nh}$ $n$ even}\\ \hline
(p,n) & ($\nu_e,\nu_o$) & $a_{1;p,0}$ & $a_{1;p,n}$ & $a_{1;p,2n}$ & $a_{1;p,3n}$ & \\ \hline
(4,$\infty$) & (1,0)
 &    2.48576 &  &  &  & \\(4,4) & (1,1)
 &    1.59799 &   -0.95498 &  &  & $O_{\rm h}$ \\
(4,2) & (2,1)
 &   -0.39950 &   -1.26332 &    0.71624 &  & $O_{\rm h}$ \\
(4,2) & (2,1)
 &    0.93216 &   -0.98258 &    1.29984 &  & \\(6,$\infty$) & (1,0)
 &   -2.98035 &  &  &  & \\(6,6) & (1,1)
 &    2.70941 &   -3.74359 &  &  & \\(6,4) & (1,1)
 &    7.45088 &   -13.9393 &  &  & $O_{\rm h}$ \\
(6,2) & (2,2)
 &    2.70941 &   -3.47040 &    0.00000 &   -1.40385 & \\(6,2) & (2,2)
 &   -12.1077 &   -4.77180 &    8.71207 &   -7.07772 & $O_{\rm h}$ \\
(6,2) & (2,2)
 &   -4.40279 &   -1.73520 &   -1.90082 &    1.16987 & \\(6,2) & (2,2)
 &    0.93136 &   -0.95436 &    1.04545 &   -1.41554 & \\(6,2) & (2,2)
 &    0.33868 &   -0.77601 &   -0.63361 &    0.52318 & $I_{\rm h}$ \\
(8,$\infty$) & (1,0)
 &    3.40498 &  &  &  & \\(8,6) & (1,1)
 &    1.83345 &   -1.24558 &  &  & \\(8,4) & (2,1)
 &   -0.22918 &   -1.48526 &    0.99999 &  & \\(8,4) & (2,1)
 &    2.97936 &   -1.12039 &    1.70705 &  & $O_{\rm h}$ \\
(10,$\infty$) & (1,0)
 &   -3.78269 &  &  &  & \\(10,10) & (1,1)
 &    7.74691 &   -13.5249 &  &  & \\(10,8) & (1,1)
 &    1.37562 &   -1.60639 &  &  & \\(10,6) & (1,1)
 &   -1.82513 &   -1.31296 &  &  & \\(10,4) & (2,1)
 &   -139.718 &   -123.111 &   -48.3395 &  & \\(10,4) & (2,1)
 &    0.99612 &   -1.00375 &   -1.19470 &  & $O_{\rm h}$ \\
(12,$\infty$) & (1,0)
 &    4.12619 &  &  &  & \\(12,10) & (1,1)
 &    1.34073 &   -1.10158 &  &  & \\(12,8) & (1,1)
 &   -1.74997 &   -2.08835 &  &  & \\(12,6) & (2,1)
 &    0.91262 &   -1.15888 &    0.62268 &  & \\(12,4) & (2,2)
 &   -5.68198 &   -8.02888 &   -0.72666 &    2.86225 & \\(12,4) & (2,2)
 &    5.08372 &   -2.41363 &   -1.98403 &    5.50958 & \\(12,4) & (2,2)
 &    5.05473 &   -1.95363 &    1.30127 &   -2.91433 & $O_{\rm h}$ \\
(12,4) & (2,2)
 &   -0.55198 &   -0.93569 &    0.89981 &   -0.56841 & $O_{\rm h}$ \\
(12,4) & (2,2)
 &   -1.64037 &   -0.45582 &   -1.99424 &    0.56834 & $O_{\rm h}$ \\
\hline
\end{tabular}
}
\end{table}

\begin{table}[t]
\assigncounter{tabcyclu}
\centerline{
\begin{tabular}{|cc|cccc|c|} \hline
\multicolumn{7}{|c|}{\bf Table \arabic{tabcyclu} Solutions of the shape equations}\\
\multicolumn{7}{|c|}{\bf  for the dihedral groups $D_{nd}$ $n$ odd}\\ \hline
(p,n) & ($\nu_e,\nu_o$) & $a_{1;p,0}$ & $a_{1;p,n}$ & $a_{1;p,2n}$ & $a_{1;p,3n}$ & \\ \hline
(4,3) & (1,1)
 &   -1.06533 &   -1.27331 &  &  & $O_{\rm h}$ \\
(6,5) & (1,1)
 &   -1.08376 &   -0.86455 &  &  & $I_{\rm h}$ \\
(6,3) & (2,1)
 &    0.60209 &   -0.91971 &   -0.55461 &  & $I_{\rm h}$ \\
(6,3) & (2,1)
 &    13.2460 &   -7.99805 &    8.38842 &  & $O_{\rm h}$ \\
(8,7) & (1,1)
 &   -1.30961 &   -1.54101 &  &  & \\(8,5) & (1,1)
 &    1.83345 &   -1.24558 &  &  & \\(8,3) & (2,1)
 &   -1.01858 &   -1.55591 &    0.54095 &  & \\(8,3) & (2,1)
 &    0.88277 &   -1.81826 &    2.21627 &  & $O_{\rm h}$ \\
(10,9) & (1,1)
 &   -1.72153 &   -1.27078 &  &  & \\(10,7) & (1,1)
 &    6.00780 &   -9.66535 &  &  & \\(10,5) & (2,1)
 &   -0.84286 &   -1.34290 &   -0.73338 &  & $I_{\rm h}$ \\
(10,3) & (2,2)
 &    2.72446 &   -7.51623 &    3.45336 &   -5.79075 & \\(10,3) & (2,2)
 &   -1.30071 &   -0.41447 &   -0.92679 &    0.90541 & $I_{\rm h}$ \\
(10,3) & (2,2)
 &   -1.57411 &   -0.19827 &    0.77110 &   -1.02866 & $O_{\rm h}$ \\
(12,11) & (1,1)
 &   -2.37941 &   -2.98770 &  &  & \\(12,9) & (1,1)
 &   -4.89423 &   -7.23640 &  &  & \\(12,5) & (2,1)
 &    3.54583 &   -1.17014 &   -0.25936 &  & \\(12,5) & (2,1)
 &    1.18220 &   -0.61091 &    0.98335 &  & $I_{\rm h}$ \\
\hline
\end{tabular}
}
\end{table}

If the z-axis is an $n$ fold rotation axis, then the shapes are invariant under
the cyclic group $C_n$. Since $p$ is even, the shapes have inversion symmetry.
Thus they are invariant under the groups $C_{nh}$ for even $n$ and under
$S_{2n}$ for odd $n$.

Only amplitudes $a_{1;p,kn}$ with integer $k$ running from $-[p/n]$ to $[p/n]$
contribute. If in particular $n>p$, then only $a_{1;p,0}$ contributes and since
$\langle p0;p0,p0\rangle$ does not vanish for even $p$, there exist for all
$p>2$ shapes invariant under any rotations around the z-axis, that is under
$C_{\infty h}$. If $n\le p$, then several amplitudes contribute. Real $r$ implies the condition $a_{1;p,-m}=(-)^m a^*_{1;p,m}$. Distinguishing
between real and imaginary part of the amplitudes one obtains $2[p/n]+1$
equations for $2[p/n]+1$ unknowns. By an appropriate rotation around the z-axis
$a_{1;p,n}$ can be made real. Then only $2[p/n]$ unknowns are left. I did not
find solutions for real $a_{1;p,n}\not=0$ and complex $a_{1;p,kn}$ for
$|k|\not=0,1$, but I cannot exclude the existence of such solutions.

If one requires that all $a_{1;p,kn}$ are real, then it is easy to find solutions.
They are invariant under dihedral groups $D_{nh}$ for even $n$, and under
$D_{nd}$
for odd $n$. Shapes were determined up to $p=12$ and $\nu:=[p/n]+1\le 4$.

The number $\nu$ of independent amplitudes is separated into
$\nu_e$ amplitudes $a_{1;p,kn}$ with even $k$ and $\nu_o$ amplitudes with odd $k$.
The shape eq. expresses the amplitude for even $k$ on the r.h.s. by products of
two amplitudes with even $k$s and products of two amplitudes with odd $k$s, but no mixed contributions, whereas
the eqs. for the amplitudes with odd $k$s contain on the r.h.s. mixed products
of one amplitude with even and one with odd $k$. As a consequence to any solution
with non-vanishing odd amplitudes there exists a second one with reversed sign
of the odd amplitudes. In the tables only one of these solutions is given.

\subsection{Tetrahedral, octahedral and icosahedral groups}

\newcommand{\cS}{{\cal S}}
\begin{table}[ht]
\assigncounter{tabted}
\centerline{
\begin{tabular}{|cc|ccccc|c|} \hline
\multicolumn{8}{|c|}{\bf Table \arabic{tabted} Solutions of the shape equations}\\
\multicolumn{8}{|c|}{\bf for the tetrahedral group $T_h$}\\ \hline
p & ($\nu_e,\nu_o$) & $a_{1;p,0}$ & $a_{1;p,2}$ & $a_{1;p,4}$ & $a_{1;p,6}$ & $a_{1;p,8}$ & \\
 & & $a_{1;p,10}$ & $a_{1;p,12}$ & $a_{1;p,14}$ & $a_{1;p,16}$ & $a_{1;p,18}$ & \\ & & $a_{1;p,20}$ & $a_{1;p,22}$ & & & & \\ \hline
6 & (1,1)
 &    0.33868 &   -0.77601 &   -0.63361 &    0.52318 &  & $I_{\rm h}$ \\
10 & (1,1)
 &    0.82311 &   -0.66870 &   -0.82942 &   -0.13114 &   -0.98720 & \\ & 
 &    0.48044 &  &  &  &  & $I_{\rm h}$ \\
12 & (2,1)
 &   -6.12511 &   -4.20630 &    9.82449 &    16.5288 &   -19.1622 & \\ & 
 &   -10.4288 &    0.69076 &  &  &  &  \\
12 & (2,1)
 &    0.83701 &   -0.23817 &    0.36961 &    0.93590 &    0.14896 & \\ & 
 &   -0.59051 &    0.49855 &  &  &  & $I_{\rm h}$ \\
16 & (2,1)
 &   -0.56063 &   -0.48811 &   -2.41712 &    1.26394 &    0.99991 & \\ & 
 &    0.49220 &    2.10476 &   -1.00183 &   -1.20704 &  & $I_{\rm h}$ \\
16 & (2,1)
 &    0.98186 &   -0.40466 &    0.07028 &    1.04785 &    0.40520 & \\ & 
 &    0.40806 &    0.54821 &   -0.83056 &    0.44283 &  &  \\
18 & (2,2)
 &    0.00360 &   -0.68341 &   -0.14607 &   -0.65318 &    0.61488 & \\ & 
 &    0.70461 &   -0.70238 &   -0.46433 &    0.20888 &    0.59331 &  \\
18 & (2,2)
 &    2.32221 &   -3.88457 &   -1.49720 &   -1.72639 &   -0.96421 & \\ & 
 &   -0.00998 &   -2.11050 &   -0.15926 &   -1.76012 &    3.00022 &  \\
18 & (2,2)
 &    0.45850 &    0.72989 &   -0.13223 &   -0.01543 &   -0.89094 & \\ & 
 &    0.68875 &    0.37789 &   -0.39434 &   -0.58803 &   -0.50005 & $I_{\rm h}$ \\
18 & (2,2)
 &   -4.73748 &    0.38021 &    2.98822 &   -4.42878 &    2.25081 & \\ & 
 &    9.29448 &    3.98375 &   -5.72487 &    3.68818 &    0.56796 &  \\
20 & (2,1)
 &    1.38678 &   -0.45717 &   -0.35519 &    0.89628 &    0.49923 & \\ & 
 &    0.59294 &    0.86536 &    0.08246 &    1.02995 &   -0.82637 & \\ & 
 &    0.40742 &  &  &  &  & $I_{\rm h}$ \\
22 & (2,2)
 &   -246.049 &   -310.255 &    115.358 &   -273.782 &    161.314 & \\ & 
 &    104.246 &    128.285 &    83.0346 &    118.399 &   -135.985 & \\ & 
 &    197.709 &    272.876 &  &  &  &  \\
22 & (2,2)
 &   -0.16701 &   -1.08554 &   -0.64291 &   -0.95793 &    2.35878 & \\ & 
 &    0.36474 &   -0.42041 &    0.29053 &   -2.02467 &   -0.47579 & \\ & 
 &    1.12764 &    0.95476 &  &  &  &  \\
22 & (2,2)
 &    2.63933 &   -9.21867 &   -1.96687 &   -4.21038 &    0.54457 & \\ & 
 &   -1.31708 &   -1.88937 &   -0.28461 &   -3.39911 &    0.30053 & \\ & 
 &   -1.11601 &    7.22239 &  &  &  &  \\
22 & (2,2)
 &    2.63933 &   -3.50688 &   -1.96687 &   -7.01919 &    0.54457 & \\ & 
 &    5.59288 &   -1.88937 &    3.69040 &   -3.39911 &   -5.87815 & \\ & 
 &   -1.11601 &    3.97000 &  &  &  & $I_{\rm h}$ \\
22 & (2,2)
 &    0.67761 &    0.69725 &   -0.19515 &   -0.34288 &   -0.82641 & \\ & 
 &    0.84351 &   -0.26707 &    0.48523 &    0.03159 &   -0.75424 & \\ & 
 &   -0.71327 &   -0.39702 &  &  &  &  \\
22 & (2,2)
 &   -1.68119 &    2.41122 &    1.93518 &   -1.18573 &   -2.47489 & \\ & 
 &    2.91702 &    1.68361 &    1.67804 &    4.15670 &   -2.60831 & \\ & 
 &   -0.22901 &   -1.37298 &  &  &  &  \\
\hline
\end{tabular}

}
\end{table}

\begin{table}[t]
\centerline{
\begin{tabular}{|cc|ccccc|c|} \hline
\multicolumn{8}{|c|}{\bf Table \arabic{tabted} cont. Solutions of the shape equations}\\
\multicolumn{8}{|c|}{\bf for the tetrahedral group $T_h$ cont.}\\ \hline
p & ($\nu_e,\nu_o$) & $a_{1;p,0}$ & $a_{2;p,2}$ &  $a_{1;p,4}$ & $a_{1;p,6}$ & $a_{1;p,8}$ & \\
 & & $a_{1;p,10}$ & $a_{1;p,12}$ & $a_{1;p,14}$ & $a_{1;p,16}$ & $a_{1;p,18}$ & \\ & & $a_{1;p,20}$ & $a_{1;p,22}$ & $a_{1;p,24}$ & $a_{1;p,26}$ & & \\ \hline 
26 & (2,2)
 &   -12.9853 &   -17.3122 &    4.24577 &   -16.3533 &    9.95326 & \\ & 
 &    1.94686 &    6.90372 &    5.42601 &    4.24205 &    1.16469 & \\ & 
 &    4.03855 &   -6.97531 &    11.0403 &    15.8632 &  &  \\
26 & (2,2)
 &    0.44976 &    3.76044 &   -3.19985 &    1.10830 &    7.15029 & \\ & 
 &    1.21251 &    1.12982 &    0.86551 &   -4.45935 &    0.16510 & \\ & 
 &   -6.08814 &   -0.94012 &    3.61602 &   -2.81306 &  & $I_{\rm h}$ \\
26 & (2,2)
 &    0.88059 &   -5.99324 &   -2.81215 &   -2.56502 &    5.52238 & \\ & 
 &   -1.39801 &    0.66375 &   -0.71139 &   -3.85344 &   -0.12650 & \\ & 
 &   -5.19225 &    0.69595 &    2.55744 &    4.69009 &  &  \\
26 & (2,2)
 &    0.89229 &    0.89158 &   -0.39047 &   -0.59796 &   -0.44156 & \\ & 
 &    0.86347 &   -0.43012 &    0.92514 &   -0.43095 &    0.18639 & \\ & 
 &   -0.46987 &   -1.08764 &   -0.62933 &   -0.44414 &  &  \\
\hline
\end{tabular}
}
\end{table}

\begin{table}[ht]
\assigncounter{tabeqshape}
\centerline{
\begin{tabular}{|c|} \hline
\bf Table \arabic{tabeqshape}: Equal shapes \\ \hline
$\cS_{D_{\infty},4} = \cS_{D_2,4,2},$ \\
$\cS_{D_4,4} = \cS_{D_2,4,1} = \cS_{D_3,4} = \cS_{O,4}.$ \\ \hline
$\cS_{D_{\infty},6} = \cS_{D_2,6,4},$ \\
$\cS_{D_6,6} = \cS_{D_2,6,1} = \cS_{D_2,6,3},$ \\
$\cS_{D_4,6} = \cS_{D_2,6,2} = \cS_{D_2,6,3} = \cS_{O,6},$ \\
$\cS_{D_2,6,5} = \cS_{D_5,6} = \cS_{D_3,6,1} = \cS_{T,6} = \cS_{I,6}.$ \\ \hline
$\cS_{D_4,8,2} = \cS_{D_3,8,2} = \cS_{O,8}.$ \\ \hline
$\cS_{D_4,10,2} = \cS_{D_3,10,3} = \cS_{O,10},$ \\
$\cS_{D_5,10} = \cS_{D_3,10,2} = \cS_{T,10} = \cS_{I,10}.$ \\ \hline
$\cS_{D_4,12,3} = \cS_{O,12,2},$ \\
$\cS_{D_4,12,5} = \cS_{O,12,3},$ \\
$\cS_{D_4,12,4} = \cS_{O,12,1},$ \\
$\cS_{D_5,12,4} = \cS_{T,12,2} = \cS_{I,12}.$ \\ \hline
\end{tabular}}
\end{table}

\begin{table}[t]
\assigncounter{taboct}
\centerline{
\begin{tabular}{|cc|ccccc|} \hline
\multicolumn{7}{|c|}{\bf Table \arabic{taboct} Solutions of the shape equations}\\
\multicolumn{7}{|c|}{\bf for the octahedral group $O_h$}\\ \hline
p & $\nu$ & $a_{1;p,0}$ & $a_{1;p,4}$ & $a_{1;p,8}$ & $a_{1;p,12}$ & $a_{1;p,16}$ \\
 & & $a_{1;p,20}$ & $a_{1;p,24}$ & & & \\ \hline
4 & 1
 &    1.59799 &    0.95498 &  &  &  \\
6 & 1
 &    7.45088 &   -13.9393 &  &  &  \\
8 & 1
 &    2.97936 &    1.12039 &    1.70705 &  &  \\
10 & 1
 &    0.99612 &   -1.00375 &   -1.19470 &  &  \\
12 & 2
 &   -0.55198 &   -0.93569 &    0.89981 &   -0.56841 &  \\
12 & 2
 &    5.05473 &    1.95363 &    1.30127 &    2.91433 &  \\
12 & 2
 &   -1.64037 &    0.45582 &   -1.99424 &   -0.56834 &  \\
14 & 1
 &    1.27193 &   -0.93532 &   -1.00368 &   -1.21871 &  \\
16 & 2
 &   -1.25843 &   -2.46818 &    0.71252 &    1.71627 &   -1.52220 \\
16 & 2
 &    6.23060 &    1.98796 &    1.77254 &    1.91038 &    3.42907 \\
16 & 2
 &   -7.22633 &    3.60347 &   -5.11674 &   -8.22622 &   -1.60500 \\
18 & 2
 &   -20.7093 &    32.0293 &   -71.4902 &    109.660 &   -11.7988 \\
18 & 2
 &   -3.80429 &    2.89479 &   -0.31593 &    5.60740 &    2.23271 \\
18 & 2
 &    1.04507 &   -0.44079 &   -1.43304 &    0.18341 &   -1.13512 \\
20 & 2
 &    1.59394 &   -0.85772 &    0.65038 &    1.28437 &    1.54183 \\ & 
 &    0.26933 &  &  &  &  \\
22 & 2
 &   -1.20555 &    0.85156 &   -0.10269 &    0.83004 &    1.41591 \\ & 
 &    0.57425 &  &  &  &  \\
22 & 2
 &   -0.22188 &   -0.74186 &    2.78358 &   -0.47953 &   -2.36217 \\ & 
 &    1.34346 &  &  &  &  \\
22 & 2
 &    2.68380 &   -0.84569 &   -3.04632 &   -1.10895 &   -0.08719 \\ & 
 &   -2.72485 &  &  &  &  \\
24 & 3
 &    1.18937 &    0.54116 &   -0.11816 &    1.28275 &   -0.82343 \\ & 
 &    0.61481 &    0.68382 &  &  &  \\
24 & 3
 &    1.67166 &    0.04356 &    0.12474 &    1.32901 &    0.00398 \\ & 
 &    1.05199 &    0.65127 &  &  &  \\
24 & 3
 &    1.35890 &   -0.55989 &    0.33293 &    0.71232 &    0.94372 \\ & 
 &    1.02056 &    0.27116 &  &  &  \\
24 & 3
 &    1.85881 &    0.00554 &    0.86643 &   -0.36989 &    1.76993 \\ & 
 &    0.50703 &    0.77945 &  &  &  \\
24 & 3
 &    0.75146 &    0.64693 &    0.61258 &   -1.06502 &    0.92200 \\ & 
 &   -0.46093 &    0.64810 &  &  &  \\
26 & 2
 &   -1.87265 &    1.19575 &    0.00293 &    0.73397 &    1.43595 \\ & 
 &    1.71925 &    0.82798 &  &  &  \\
26 & 2
 &   -0.08100 &   -3.16660 &    7.90156 &    1.47491 &   -4.48413 \\ & 
 &   -6.19642 &    4.25103 &  &  &  \\
26 & 2
 &    4.63643 &   -0.57937 &   -5.85327 &   -2.88496 &   -0.19160 \\ & 
 &    0.38293 &   -5.16865 &  &  &  \\
\hline
\end{tabular}

}
\end{table}

\begin{table}[t]
\centerline{
\begin{tabular}{|cc|ccccc|} \hline
\multicolumn{7}{|c|}{\bf Table \arabic{taboct} cont. Solutions of the shape equations}\\
\multicolumn{7}{|c|}{\bf for the octahedral group $O_h$ }\\ \hline
p & $\nu$ & $a_{1;p,0}$ & $a_{1;p,4}$ & $a_{1;p,8}$ & $a_{1;p,12}$ & $a_{1;p,16}$ \\
 & & $a_{1;p,20}$ & $a_{1;p,24}$ & $a_{1;p,28}$ & $a_{1;p,32}$ & \\ \hline
28 & 3
 &   -11.2790 &   -15.3654 &   -2.10025 &   -3.08484 &    3.01089 \\ & 
 &    7.52651 &    1.63605 &   -11.7977 &  &  \\
28 & 3
 &   -8.19362 &   -77.0755 &   -6.62562 &    9.30006 &    32.7654 \\ & 
 &    48.2929 &    35.2893 &   -41.3066 &  &  \\
28 & 3
 &    1.31496 &   -0.12505 &   -0.16182 &    1.13151 &    0.54722 \\ & 
 &   -0.02126 &    1.04212 &    0.39059 &  &  \\
28 & 3
 &    4.96143 &    2.93182 &    3.87509 &   -3.45602 &    0.33561 \\ & 
 &    4.06348 &   -1.77305 &    3.70403 &  &  \\
28 & 3
 &    8.08355 &    2.16990 &    4.11477 &   -1.80224 &    1.82715 \\ & 
 &    5.30833 &    0.32544 &    4.48484 &  &  \\
30 & 3
 &   -10.3305 &    3.75338 &    5.18714 &    4.88790 &    4.09221 \\ & 
 &    4.37127 &    4.62163 &    7.22909 &  &  \\
30 & 3
 &   -4.12966 &    1.90227 &    1.61647 &   -0.22958 &    5.29145 \\ & 
 &   -0.70240 &    3.63614 &    2.20732 &  &  \\
30 & 3
 &   -0.32893 &   -0.47577 &    1.23931 &    1.34217 &   -1.44945 \\ & 
 &    0.41590 &   -1.13186 &    1.04467 &  &  \\
30 & 3
 &    0.80723 &    0.36595 &   -1.83491 &   -0.45764 &   -0.88932 \\ & 
 &    1.37893 &    0.22661 &   -1.34800 &  &  \\
30 & 3
 &    73.5784 &   -105.971 &    206.495 &   -395.192 &    731.230 \\ & 
 &   -842.698 &    143.693 &    7.16578 &  &  \\
32 & 3
 &   -0.94505 &   -1.62873 &    1.80496 &   -2.69679 &   -1.24200 \\ & 
 &    1.76241 &    2.78195 &   -1.70814 &   -0.86640 &  \\
32 & 3
 &    2.02108 &   -0.08638 &   -0.29673 &    1.28426 &    1.07854 \\ & 
 &    0.37970 &    0.19005 &    1.46548 &    0.62050 &  \\
32 & 3
 &   -1.16350 &    1.20893 &   -0.44735 &    0.08591 &   -0.46335 \\ & 
 &   -1.25548 &   -1.57057 &   -0.62832 &    0.12415 &  \\
32 & 3
 &    7.53901 &    12.2486 &    8.93730 &   -5.85501 &   -5.29201 \\ & 
 &   -0.18349 &    1.41254 &   -7.58121 &    10.0563 &  \\
32 & 3
 &    47.6029 &    30.0645 &    24.1029 &   -3.30451 &   -1.96647 \\ & 
 &    7.99337 &    11.6529 &   -3.80089 &    35.2026 &  \\
34 & 3
 &   -11.4914 &   -9.42515 &    25.0654 &    30.3572 &   -9.41248 \\ & 
 &   -9.23599 &    1.92303 &   -17.1336 &    25.2407 &  \\
34 & 3
 &    4.10996 &   -1.74467 &   -1.65191 &    0.50294 &   -3.74252 \\ & 
 &   -2.26974 &    0.11285 &   -3.55783 &   -2.00760 &  \\
34 & 3
 &   -5.79444 &    0.80954 &    4.16799 &    5.11033 &    0.56450 \\ & 
 &    0.94328 &    2.59973 &    0.35243 &    5.36668 &  \\
34 & 3
 &    0.51337 &    0.24287 &   -1.08014 &   -0.06906 &   -0.82773 \\ & 
 &    0.02861 &    0.97644 &   -0.20870 &   -0.77065 &  \\
34 & 3
 &    1.09993 &   -0.90039 &    0.44903 &   -0.02780 &   -0.34114 \\ & 
 &   -0.84004 &   -1.20752 &   -0.96928 &   -0.08435 &  \\
\hline
\end{tabular}
}
\end{table}

\begin{table}[t]
\assigncounter{tabdod}
\centerline{
\begin{tabular}{|cc|ccccc|} \hline
\multicolumn{7}{|c|}{\bf Table \arabic{tabdod} Solutions of the shape equations}\\
\multicolumn{7}{|c|}{\bf for the icosahedral group $I_h$}\\ \hline
p & $\nu$ & $a_{1;p,0}$ & $a_{1;p,5}$ & $a_{1;p,10}$ &$a_{1;p,15}$ & $a_{1;p,20}$ \\
 & & $a_{1;p,25}$ & $a_{1;p,30}$ & $a_{1;p,35}$ & & \\ \hline
6 & 1
 &   -1.08376 &    0.86454 &  &  &  \\
10 & 1
 &   -0.84286 &   -1.34290 &   -0.73338 &  &  \\
12 & 1
 &    1.18220 &   -0.61091 &    0.98335 &  &  \\
16 & 1
 &    2.53390 &    1.87102 &   -2.40430 &   -1.86451 &  \\
18 & 1
 &   -1.24551 &    0.70244 &   -0.56268 &    1.11003 &  \\
20 & 1
 &    0.35816 &    1.06413 &    1.72730 &    0.53368 &    0.94937 \\
22 & 1
 &   -9.42141 &   -3.84903 &    6.02057 &   -6.60823 &   -6.45637 \\
24 & 1
 &    1.28346 &   -0.84415 &    0.57424 &   -0.59346 &    1.25220 \\
26 & 1
 &   -3.35369 &   -6.46073 &   -4.17676 &    7.32138 &    3.70065 \\ & 
 &    5.20342 &  &  &  &  \\
28 & 1
 &   -21.2790 &   -4.33343 &    10.4040 &   -11.4419 &    12.2013 \\ & 
 &    14.1643 &  &  &  &  \\
30 & 2
 &   -1.29745 &    1.00742 &   -0.67877 &    0.53443 &   -0.66788 \\ & 
 &    1.41535 &   -0.00751 &  &  &  \\
30 & 2
 &    0.43597 &   -0.78987 &   -0.90073 &   -1.69849 &   -0.33963 \\ & 
 &   -0.64517 &   -0.90325 &  &  &  \\
30 & 2
 &    3.94227 &   -1.15076 &    6.83983 &    4.80458 &    4.41657 \\ & 
 &   -3.58280 &    3.85631 &  &  &  \\
32 & 1
 &   -25.3074 &   -34.3654 &   -6.83507 &    30.6595 &   -30.5265 \\ & 
 &   -22.3716 &   -28.6144 &  &  &  \\
34 & 1
 &    7.10336 &    0.32668 &   -2.79493 &    3.20114 &   -3.38581 \\ & 
 &    3.53378 &    4.72435 &  &  &  \\
36 & 2
 &   -1.07451 &    2.68870 &    1.75129 &    1.63551 &   -3.30158 \\ & 
 &   -0.87907 &   -2.15745 &   -2.00621 &  &  \\
36 & 2
 &   -0.83679 &   -0.11942 &   -1.82417 &   -0.21073 &    1.13536 \\ & 
 &    1.30552 &   -0.61666 &    1.06790 &  &  \\
36 & 2
 &    1.28515 &   -1.18256 &    0.83404 &   -0.59249 &    0.54385 \\ & 
 &   -0.77680 &    1.60342 &   -0.01678 &  &  \\
\hline
\end{tabular}
}
\end{table}

Other subgroups of the O(3) are the tetrahedral, the octahedral and the
icosahedral groups. Due to inversion symmetry the corresponding groups are
$T_h$, $O_h$, and $I_h$. Only floating bodies of equilibrium with mirror symmetry have been found, since all these groups have mirror planes.
I do not know whether there exist any shapes without this symmetry.

Solutions of the shape
equations are listed in the following tables. For the tetrahedral group
one can again distinguish between even and odd contributions.
The numbers $\nu(p)$ of linearly independent functions $r_K$ obey
\bea
\nu_{\rm oct}(p) &=& \nu_{\rm tet\,e}(p) = \nu_{\rm tet\,o}(p+6), \\
\nu_{\rm oct}(p+12) = \nu_{\rm oct}(p)+1, &&
\nu_{\rm ico}(p+30) = \nu_{\rm ico}(p)+1.
\eea

The spherical harmonics invariant under these groups are constructed by
requiring that they are invariant under two elements generating the group. The first group
element is a rotation about the z-axis by $\pi$ for $T_h$, $\pi/2$ for $O_h$,
and $2\pi/5$ for $I_h$. The second is a rotation about another axis by $2\pi/3$, $\pi/2$, and $2\pi/5$ for $T_h$, $O_h$, and $I_h$, resp., described by the rotation matrix
$D(\alpha,\beta,\gamma)$,

\be
\begin{array}{ll}
{\rm tetrahedral} & D(\frac{\pi}2,\frac{\pi}2,0), \\
{\rm octahedral} & D(0,\frac{\pi}2,0), \\
{\rm icosahedral} & D(0,\beta,\frac{\pi}5),
\,\,\cos\beta=\frac 1{\sqrt 5},\,\,\sin\beta=\frac 2{\sqrt 5}.
\end{array}
\ee

If for $T_h$ the amplitudes $a_{1;p,2k}$ for odd $k$ vanish, then the solutions
belong to
$O_h$. I did not tabulate them in the table for the tetrahedral group. If one
wishes to insert them, then one should multiply the amplitudes $a_{4k}$ by
$(-)^k$, since different orientations of the equivalent mirror-planes have been chosen for $T_h$ and $O_h$. For
$O_h$ and $I_h$ one can no longer distinguish between even and odd amplitudes.

The solutions which correspond to the groups $O_h$ and $I_h$ are indicated in
the tables for the dihedral groups. The solutions invariant under $I_h$ are
indicated in the table for the tetrahedral group. Some solutions
appear several times in the tables in different orientations.
A list of equal shapes is given in table \arabic{tabeqshape}.
The shapes are denoted by
${\cS}_{G,p,i}$, where $G$ denotes one of the groups $D_n$,
$T$, $O$ and $I$, resp. and $i$ numbers the solutions for given $G$ and $p$
($i$ is omitted, if only one solution for $G$ and $p$ is listed).

Maple was able to solve the set of shape equations without giving any further
hint apart from the case $(\nu_e,\nu_o)=(2,2)$ for the tetrahedral group.
Therefore in this case the equations were rearranged as shown in appendix \ref{c22}.
Apparently the case $(2,2)$ is easier solved for the dihedral groups, since some of the
coefficients $w_{j,i_1,i_2}$ vanish in this later case.

In general $2^{\nu}$ solutions were obtained including the trivial one, for which
all amplitudes vanish. Only real solutions are listed, since in the calculation $a_{1;p,-m}$ was expressed by $(-)^m a_{1;p,m}$, whereas it should be 
$(-)^m a^*_{1;p,m}$.

\paragraph{Case $\nu=2$} As an example the set of equations (\ref{alphaeq}) for the case $\nu=2$ is considered,
\bea
\alpha_1 &=& w_{111} \alpha_1^2+2 w_{112} \alpha_1\alpha_2 + w_{122} \alpha_2^2,
\\
\alpha_2 &=& w_{112} \alpha_1^2+2 w_{122} \alpha_1\alpha_2 + w_{222} \alpha_2^2.
\eea
The ratio
\be
r=\frac{\alpha_2}{\alpha_1},
\ee
is introduced for the non-trivial solutions. Then the equation
\be
r=\frac{w_{112}+2w_{122}r+w_{222}r^2}{w_{111}+2w_{112}r+w_{122}r^2}
\ee
is obtained, which rewrites as an equation of third order in $r$,
\be
w_{122} r^3 + (2w_{112}-w_{222}) r^2 + (w_{111}-2w_{122}) r - w_{112} = 0.
\label{eqr}
\ee
Thus there are three non-trivial solutions. One or three of them are real.

\section{Higher Orders in the deformation}

As in the preceeding section only central symmetric bodies are considered. Then the
expansion of $r$ and $h$ in spherical harmonics contains only those with even
$l$.
In higher orders the equation for $V_{{\rm a}n}$ can be used to determine $h_n(\beta,\alpha)$ up to an $\beta,\alpha$-independent contribution $h_{n;00}$.
The difference $Z_{{\rm a}n}-\ct r_0 V_{{\rm a}n}$ yields the condition
\be
A_{1l}r_0a_{n,lm} + A_{2l}(h_{1;00}a_{n-1,lm}+h_{n-1;00}a_{1,lm})
+ 2A_{3l} (a_1 a_{n-1})_{;lm}= I_{n;lm}
\ee
for $l>0$, where $I_{n;lm}$ contains only terms $a_m$ with $m<n-1$. For
$l\not=p$ it yields
\be
a_{n,lm} = -\frac 1{A_{1l}} \left(I_{n;lm}-2A_{3l} (a_1 a_{n-1})_{;lm}\right).
\ee
For $l=p$ one obtains a set of linear eqs. for the $a_{n-1,pm}$
\be
a_{n-1;pm} -2\sum_{m'} M_{mm'} a_{n-1;pm'} = \hat I_{n;m}
\ee
with
\bea
M_{mm'} &=& \langle pm;pm-m',pm' \rangle \frac{a_{1;pm-m'}}{\gamma}, \\
\hat I_{n;m} &=& -\frac{h_{n-1;00}}{h_{1;00}}
+ \frac{I_{n;pm}}{s_0h_{1;00}P_{p,-1}(\ct)} \\ 
&+&\frac{c_0s_0+2P_{p,-1}(\ct)}{s_0h_{1;00}} \sum_{l\not=p,m_1m_2}
\langle p0;p0,l0 \rangle \langle pm;pm_1,lm_2 \rangle a_{1;pm_1} a_{n-1;lm_2}.
\nonumber
\eea
This set of equations has a unique solution, if the matrix $M$ does not have an
eigen-value $1/2$. An eigen-value $1/2$ appears, if the shape-eqs. have a double-solution.
This means that besides $a_{1;p.}$ also $a_{1;p.}+\delta a_{1;p.}$ is solution to first order in $\delta a_{1;p.}$, which yields the equations
\be
\delta a_{1;pm} -2\sum_{m'} M_{mm'} \delta a_{1;pm'} = 0.
\ee
This is the case, if one does not fix the orientation of the body, since then
the change of $a_{1;p.}$ by infinitesimal rotations yields such solutions
$a_{1;p.}+\delta a_{1;p.}$. Here we have fixed the symmetry axes in the calculations. Thus this case does not appear. We cannot exclude that in special cases a
double-solution appears. We have determined the eigen-values of $M$ for all shapes
listed in the tables. No eigenvalue $1/2$ appeared. The closest one were 0.4541
for $\cS_{D_4,12,5}$ and 0.5979 for $\cS_{T,26,2}$. Always one of
the eigenvalues equals 1 for the eigen-vector $a_{1;p.}$.

\subsection{Example for double-solution}

We give an example for a double-solution by further considering the case
$\nu=2$.
Starting from eq. (\ref{eqr}) and
denoting the solutions by $r_1$, $r_2$, $r_3$ one obtains
\bea
w_{111} &=& w_{122} (2+r_1r_2+r_1r_3+r_2r_3), \\
w_{112} &=& w_{122} r_1r_2r_3, \\
w_{222} &=& w_{122} (r_1+r_2+r_3+2r_1r_2r_3).
\eea
Considering the solution $r=r_1$ one obtains
\bea
\alpha_1=\frac 1{w_{122}N}, && \alpha_2=\frac{r_1}{w_{122}N}, \nn
N &=& 2+r_1r_2+r_1r_3+r_2r_3+2r_1^2r_2r_3+r_1^2.
\eea
This yields the matrix $M$ with the matrix-elements
\bea
M_{ij} &=& w_{ij1} \alpha_1 + w_{ij2} \alpha_2, \\
M_{11} &=& \frac{2+r_1r_2+r_1r_3+r_2r_3+r_1^2r_2r_3}N, \\
M_{12} = M_{21} &=& \frac{r_1(1+r_2r_3)}N, \\
M_{22} &=& \frac{1+r_1(r_1+r_2+r_3)+2r_1^2r_2r_3}N.
\eea
The eigenvalues of this matrix are
\be
\mu_1 =1, \quad \mu_2 = \frac{1+r_1r_2+r_1r_3+r_1^2r_2r_3}N.
\ee
One finds
\be
\mu_2-\frac 12 = \frac{(r_1-r_2)(r_1-r_3)}{2N}.
\ee
Thus one obtains an eigenvalue $\mu=1/2$ if and only if $r_1$ equals one of the other solutions $r_i$.

\subsection{Reparametrization}

By iterating the expansion in $\epsilon$ it became apparent that the
coefficients $r_{n;00}$ and $h_{n;00}$ were undetermined, whereas all other
coefficients $r_{n;lm}$ and $h_{n;lm}$ were determined.

The reason is the following: One is free to reparametrize $r$ by introducing a
new $\hat r=r s_r(\epsilon)$ with a function $s_r(\epsilon)$ that has a Taylor
expansion in $\epsilon$. Simultaneously $h$ has to be changed to $\hat h=h
s_r(\epsilon)$. Thus one is free to choose $r_{n;00}$ arbitrarily. One may for
example choose $r_{n;00}=0$ for $n>0$ or require a volume independent of
$\epsilon$.

If the $r_{n;00}$s are fixed one is still free to reparametrize $h$ to $\hat h=h
s_h(\epsilon)$, where again $s_h(\epsilon)$ must be Taylor expandable. A
possible choice is $h_{n;00}=0$ for $n>1$.

\paragraph{Acknowledgment} I am indebted to Christian Wegner for useful discussions on section 3.

\begin{appendix}

\section{Supplement to section \ref{rhohalf}\label{Gegenbauer}}

\subsection{Ultraspherical (Gegenbauer) polynomials}

For more information on ultraspherical polynomials see
\cite{Abramowitz,Bateman}.

Introducing coordinates
\be
u_i = \cos\theta_i \prod_{k=i+1}^d \sin\theta_k,
\ee
where $\theta_k$ runs from 0 to $\pi$ for $k>1$ and $\theta_1=0,\pi$ allows to
write the ul\-tra\-sphe\-ri\-cal harmonics as
\bea
&& Y^{(d)}_{m_d,m_{d-1},...,m_2,m_1}(\theta_d,...\theta_1) \nn
&=& \sin^{m_{d-1}}\theta_d
\frac{G^{(m_{d-1}+d/2-1)}_{m_d-m_{d-1}}(\cos\theta_d)}
{\sqrt{N^{(m_{d-1}+d/2-1)}_{m_d-m_{d-1}}}}
Y^{(d-1)}_{m_{d-1},...,m_2,m_1}(\theta_{d-1},...\theta_1), \\
&& Y^{(1)}_{m_1}(\theta_1) = \frac 1{\sqrt 2} \cos^{m_1}{\theta_1}
\eea
with the restriction $m_k\ge m_{k-1}$ and $m_1=0,1$. For given $d$ and $m_d$
these are
$(2m_d+d-2)(m_d+d-3)!/((d-2)!m_d!)$ ultraspherical harmonics. They form a
complete orthonormal set of functions and obey eq. (\ref{Lapld}), where $n=m_d$
and $m$ stands for the set $\{m_{d-1},...,m_1\}$.

Gegenbauer's addition theorem \cite{Gegenbauer} reads
\bea
&& (n+\alpha) \Gamma(\alpha) G^{(\alpha)}_n
\left(xx'+\sqrt{(1-x^2)(1-x^{\prime 2})}\cos\phi\right) \nn
&=&\sum_{l=0}^n
\frac{(1-x^2)^{l/2}(1-x^{\prime 2})^{l/2}}{N^{(\alpha+l)}_{n-l}}
G^{(\alpha+l)}_{n-l}(x)G^{(\alpha+l)}_{n-l}(x') \nn
&\times& (l+\alpha-1/2) \Gamma(\alpha-1/2) G^{(\alpha-1/2)}_l(\cos\phi),
\label{addC}
\eea
with the norm
\bea
\int_{-1}^{+1} \de x(1-x^2)^{\alpha-1/2} G^{(\alpha)}_n(x) G^{(\alpha)}_m(x)
= N^{(\alpha)}_n \delta_{n,m}, \label{orthC} \\
N^{(\alpha)}_n =
\frac{2^{1-2\alpha}\pi\Gamma(n+2\alpha)}{n!(n+\alpha)\Gamma^2(\alpha)}.
\label{normC}
\eea
Repeated use of this addition theorem with $\w u=x\w e_d+\sqrt{1-x^2}\w u'$,
$\w e_d$ unit vector, $x=\cos\theta_d(\w u)$, similarly for $\w v$, and
$\w u'\cdot\w v'=\cos\phi$ yields eq. (\ref{addY}).

\subsection{Determination of the expansion coefficients $c_n$\label{calcz}}

The Legendre polynomials are ultraspherical polynomials,
$P_n(x)=G^{(1/2)}_n(x)$. Thus it is sufficient to consider the expansion
eq. (\ref{expz}). Orthogonality of the polynomials (\ref{orthC}) yields
\be
c_n N_{2n}^{(d/2-1)} = 2\int_0^1 \de x\, x (1-x^2)^{d/2-3/2} G^{(d/2-1)}_{2n}(x).
\ee
$G$ is represented by Rodriguez' formula
\bea
G^{(d/2-1)}_{2n}(x) &=& \kappa_{2n}^{(d/2-1)} (1-x^2)^{3/2-d/2}
\frac{\de^{2n}}{\de x^{2n}} (1-x^2)^{d/2-3/2+2n}, \\
\kappa_{2n}^{(d/2-1)} &=& \frac{\Gamma(\frac{d-1}2)\Gamma(2n+d-2)} {2^{2n}(2n)!\Gamma(d-2)\Gamma(2n+\frac{d-1}2)}.
\eea
Thus
\be
c_n N_{2n}^{(d/2-1)}= 2\kappa_{2n}^{(d/2-1)} \int_0^1 \de x\, x \frac{\de^{2n}}{\de x^{2n}} (1-x^2)^{d/2-3/2+2n}.
\ee
Partial integration yields \cite{Falconer}
\be
2\frac{\kappa_{2n}^{(d/2-1)}}{\kappa_{2n-2}^{(d/2+1)}} G^{(d/2+1)}_{2n-2}(0).
\ee
With
\be
G^{(d/2+1)}_{2n-2}(0) = \left({n+d/2-1 \atop n-1}\right)
\ee
one obtains the expression for $c_n$ in eq.(\ref{expz}).

\section{The case $(\nu_{\rm e},\nu_{\rm o})=(2,2)$\label{c22}}

The shape-equations can be written
\bea
0 &=& c_x - c'_{xuv}, \label{nullx} \\
0 &=& c_y - c'_{yuv}, \label{nully} \\
0 &=& c_{uu} u + c_{uv} v, \label{nullu} \\
0 &=& c_{uv} u + c_{vv} v. \label{nullv}
\eea
with
\bea
c_x &=& x-w_{000}x^2-2w_{002}xy-w_{022}y^2, \\
c_y &=& y-w_{002}x^2-2w_{022}xy-w_{222}y^2, \\
c'_{xuv} &=& w_{011}u^2+2w_{013}uv+w_{033}v^2, \label{csxuv} \\
c'_{yuv} &=& w_{211}u^2+2w_{213}uv+w_{233}v^2, \label{csyuv} \\
c_{uu} &=& 1-2w_{011}x-2w_{211}y, \\
c_{uv} &=& -2w_{013}x-2w_{213}y, \\
c_{vv} &=& 1-2w_{033}x-2w_{233}y.
\eea
Eqs. (\ref{nullu}, \ref{nullv}) yield
\bea
c_{uu}c_{vv}-c^2_{uv}=0. \label{nulluv1}, \\
\frac uv = - \frac{c_{uv}}{c_{uu}} = - \frac{c_{vv}}{c_{uv}}. \label{uv}
\eea
Insertion in (\ref{csxuv},\ref{csyuv}) yields
\bea
c'_{xuv} = -\frac{uv}{c_{uv}} c_{xuv}, &&
c_{xuv} = w_{011}c_{vv}-2w_{013}c_{uv}+w_{033}c_{uu}, \label{cxuv}\\
c'_{yuv} = -\frac{uv}{c_{uv}} c_{yuv}, &&
c_{yuv} = w_{211}c_{vv}-2w_{213}c_{uv}+w_{233}c_{uu}.
\eea
and substitution in eqs. (\ref{nullx},\ref{nully}) and elimination of
$\frac{uv}{c_{uv}}$ yields
\be
c_x c_{yuv} - c_y c_{xuv} = 0. \label{nullxy1}
\ee
The eqs. (\ref{nulluv1}, \ref{nullxy1}) allow the determination of
$x$ and $y$.

From (\ref{nullx},\ref{cxuv},\ref{uv}) one obtains
\be
c_x c_{vv} - c_{xuv} u^2 = 0
\ee
which allows the calculation of $u$. Finally eq. (\ref{nullu}) yields $v$.
One notes that the solutions with $u=v=0$ have been eliminated by use of
(\ref{nulluv1}, \ref{uv}). They can be obtainded from $c_x=c_y=0$.

\section{Denominator of the coefficient $\gamma$\label{dengamma}}

The Legendre functions obeys the recursion relation
\be
(p+2) P_{p+1,-1}(x) - (2p+1) xP_{p,-1}(x) + (p-1) P_{p-1,-1}(x) = 0.
\ee
Using it also for $p\rightarrow p\pm1$, one can eliminate $P_{p\pm1,-1}(x)$ and
obtain the recursion relation
\bea
&& (p+2)(p+3)(2p-1) P_{p+2,-1}(x) \nn
&+& (2p+1)\big(2p^2+2p-3-(2p-1)(2p+3)x^2\big) P_{p,-1}(x) \nn
&+&(p-1)(p-2)(2p+3) P_{p-2,-1}(x) = 0.
\eea
Substituting
\be
\frac 12 \st\ct + P_{p,-1}(\ct) = \st\ct(1-\ct^2) K_p(\ct)
\ee
one obtains with $\ct=x$
\bea
&&(p+2)(p+3)(2p-1) K_{p+2}(x) \nn
&+& (2p+1)\big(2p^2+2p-3-(2p-1)(2p+3)x^2\big) K_p(x) \nn
&+& (p-1)(p-2)(2p+3) K_{p-2}(x) \nn 
&=& \frac 12 (2p-1)(2p+1)(2p+3). \label{recK}
\eea
One easily calculates
\be
K_2=0, \quad K_4 = \frac 78.
\ee
The recursion relation (\ref{recK}) yields
\bea
K_6 &=& \frac 3{16}(1+11x^2), \nn
K_8 &=& \frac{11}{128}(9-26x^2+65x^4), \nn
K_{10} &=& \frac{13}{256} (5+89x^2-289x^4+323x^6), \nn
K_{12} &=& \frac 1{1024} (743-5032x^2+34238x^4-72353x^6+52003x^8).
\eea
Numerical calculations yield no real zeroes of $K_p(x)$ up to $p=36$. Probably
all $K_p(x)$ are strictly positive for real $x$.

\end{appendix}

\end{document}